\documentclass[aps,amssymb,showkeys]{revtex4}
\usepackage{graphicx}
\usepackage{epstopdf}
 \usepackage[bookmarks=false]{hyperref}   
\usepackage{amsmath}

\newcommand{\be}{\begin{equation}}
\newcommand{\ee}{\end{equation}}
\newcommand{\bea}{\begin{eqnarray}}
\newcommand{\eea}{\end{eqnarray}}
 
\newcommand{\ba}[1]{\begin{array}{#1}}
\newcommand{\ea}{\end{array}}

\def\G{\Gamma}
\def\g{\gamma}

\def\e{\varepsilon} 
\renewcommand{\l}{\lambda}

\newcommand{\degree}{^{ \circ}}

\begin{document}

\title{Factorization of percolation density correlation functions for clusters touching the sides of a rectangle}

\author{J. J. H. Simmons}
\email{j.simmons1@physics.ox.ac.uk}
\affiliation{Rudolf Peierls Centre for Theoretical Physics, 1 Keble Road, Oxford OX1 3NP, UK}

\author{Robert M. Ziff}
\email{rziff@engin.umich.edu}
\affiliation{Michigan Center for Theoretical Physics and Department of Chemical Engineering, University of Michigan, Ann Arbor MI 48109-2136}

 \author{Peter Kleban}
\email{kleban@maine.edu}
\affiliation{LASST and Department of Physics \& Astronomy,
University of Maine, Orono, ME 04469, USA}

\date{\today}
\begin{abstract}
In this paper we consider the density, at a point $z = x + i y$,  of critical percolation clusters that touch the left ($P_L(z)$), right ($P_R(z)$), or both ($P_{L R}(z)$) sides of a rectangular system, with open boundary conditions on the top and bottom sides.  While each of these quantities is nonuniversal and indeed vanishes in the continuum limit,  the ratio $C(z) = P_{L R}(z) / \sqrt{P_L(z) P_R(z) \Pi_h}$, where $\Pi_h$ is the probability of left-right crossing given by Cardy, is a universal function of $z$.   With wired (fixed) boundary conditions on the left- and right-hand sides, high-precision numerical simulations and theoretical arguments show that $C(z)$ goes to a constant $C_0 = 2^{7/2}\;3^{-3/4}\,\pi^{5/2}\;\G(1/3)^{-9/2} = 1.0299268\ldots$ for points far from the ends, and varies by no more than a few percent for all $z$ values.  Thus $P_{L R}(z)$ factorizes over the entire rectangle to very good approximation.  In addition, the numerical observation that $C(z)$ depends upon $x$ but not upon $y$ leads to an explicit expression for $C(z)$ via conformal field theory for a long rectangle (semi-infinite strip).   We also derive explict expressions for $P_L(z)$, $P_R(z)$, and $P_{L R}(z)$ in this geometry, first by assuming $y$-independence and then by a full analysis that obtains these quantities exactly with no assumption on the $y$ behavior. In this geometry we obtain, in addition,  the corresponding quantities in the case of open boundary conditions, which allows us to calculate $C(z)$ in the open system.  We give some theoretical results for  an arbitrary rectangle as well.  Our results also enable calculation of the finite-size corrections to the factorization near an isolated anchor point, for the case of clusters anchored at points. Finally, we present numerical results for a rectangle with periodic b.c.\ in the horizontal direction, and find $C(z)$ approaches a constant value $C_1 \approx 1.022$.
\end{abstract}

\keywords{percolation, correlation functions, factorization, conformal field theory, numerical simulation}
\maketitle

\section{Introduction} \label{intro}

Percolation at the critical point has many well-known universal properties, including universal critical exponents, scaling functions, and amplitude ratios. Universality means that the properties are the same for all realizations of the system (with a given dimensionality) in the continuum or field-theory limit.  Crossing probabilities are also universal and have received a great deal of attention since the work of Cardy \cite{Cardy92} and Langlands et al.\ \cite{LanglandsEtAl92}, and renewed interest more recently with the development of Schramm-Loewner Evolution (SLE) \cite{LawlerSchrammWerner01} and a new set of results for percolation \cite{SimmonsKlebanZiffJPA07}.

A more detailed picture of the critical system can be obtained by examining the correlations within clusters of connected sites.
In previous work \cite{KlebanSimmonsZiff06,SimmonsKlebanZiffPRE07}, we demonstrated, by use of conformal field theory and high-precision simulation,  certain exact and universal factorizations of higher-order correlation functions in terms of lower-order correlation functions for percolation clusters in two dimensions at the percolation point.  In that work, the correlation functions involved the density of critical percolation clusters constrained to touch one or two isolated boundary points, or single boundary intervals.  Here we extend those results by considering densities constrained to touch one or two distinct boundary intervals, which is a more difficult problem.

Specifically, we consider the quantities $P_L(z)$, $P_R(z)$, and $P_{L R}(z)$, which give the density of percolation clusters at a point $z=x+iy$ that touch the left, right, or both sides of a rectangle, respectively, as well as $\Pi_h$ the probability of a horizontal crossing (i.e., one or more clusters that touch both left and right sides) which is given by Cardy's formula \cite{Cardy92}.  $P_L(z)$, $P_R(z)$, and $P_{L R}(z)$ also determine the \emph{probabilities} that the given boundaries are connected to $z$, or more precisely to a disk of radius $\e$ around $z$.  Individually they are non-universal and furthermore go to zero as the lattice mesh size (or $\e$) goes to zero.  However, we find numerically and prove via conformal field theory that the ratio
\be \label{Cratio}
C(z) = \frac{P_{L R}(z) }{ \sqrt{P_L(z) P_R(z) \Pi_h} }   \; ,
\ee
is a universal function of $z$ in the limit that $\e$ goes to zero, depending only upon the boundary conditions on the sides of the rectangle.  

For most of this paper (except the last section)  the boundary conditions (b.c) are assumed to be open, or free, on the top and bottom sides, and either open or wired on the left and right sides.  The behavior of $C(z)$ near the left and right sides depends strongly on our choice of b.c.  However, for rectangles with width $W$ greater than a few times their height $H$, $C(z)$ goes exponentially to a universal constant, $C_0$, for points that are on the order of $W$ away from the left and right sides, regardless of the boundary conditions on those sides.   We find that the asymptotic value $C_0$ is the same as that found for the case of point anchors \cite{KlebanSimmonsZiff06,SimmonsKlebanZiffPRE07}:
\be \label{C0value}
C_0  = \frac{2^{7/2}\;\pi^{5/2}}{3^{3/4}\;\G(1/3)^{9/2}} = 1.0299268 \ldots \;.  
\ee
This agreement is expected because for points $z$ far from either vertical side the difference between anchoring to boundary points or small intervals becomes negligible.  

Furthermore, we find the surprising result that with wired boundary conditions on the left- and right-hand sides $C(z)$ depends only upon the horizontal coordinate $x$ and not on $y$, even though the individual functions $P_L(z)$ etc.\ have a strong dependence on $y$.  With wired boundary conditions all clusters touching the boundary are assumed to be connected together, so that if there is a crossing cluster all other clusters touching either boundary are also part of it.  For this boundary condition, $C(z)$ goes to 1 as $z$ approaches the left or right sides, and remains within a few percent of $1$ for all $z$, so that factorization is a good approximation everywhere in the rectangle.

The results simplify particularly nicely for the case of a long rectangular system which we approximate as a semi-infinite strip (of unit width).  In that case, 
we find 
\begin{eqnarray}
C(x) &=& C_0\frac{{}_2F_1(  -1/2,-1/3,7/6, e^{-2 \pi x} )}{\sqrt{{}_2F_1( -1/2,-2/3,5/6  , e^{-2 \pi x})}}  \nonumber \\
&\sim& C_0\left(1-\frac{2}{35} e^{-2 \pi x}+\frac{834}{25025} e^{-4 \pi x} + \ldots \right)   \qquad \hbox{(for $x \to \infty$)}
\end{eqnarray}
where $x$ is the distance from one end.  We also provide numerical confirmation of this result. 

\begin{figure}[htbp]
\begin{center}
\includegraphics[height=7in]{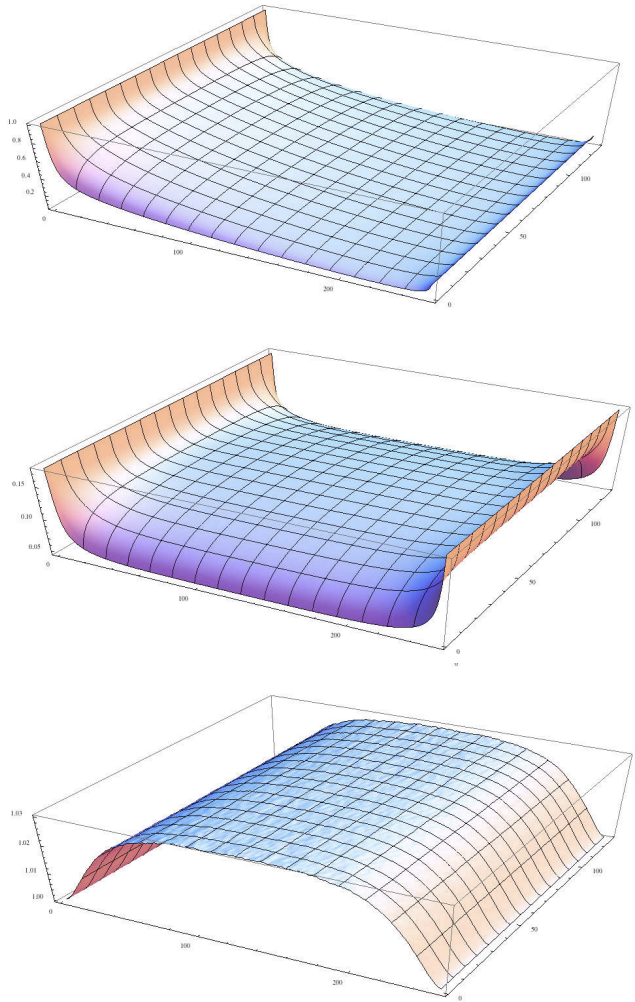}
\caption{Plots of $P_L(z)$ (top), $P_{L R}(z)$ (center), and $C(z)$ (bottom) for a $127 \times 255$ system
with wired b.c.\ on the left and right-hand sides.}
\label{wiredfig}
\end{center}
\end{figure}

If we consider open boundary conditions on the left- and right-hand sides, then  $C(z)$ remains universal and approaches $C_0$ far from the vertical sides.  However, near the left- and right-hand sides, $C(z)$ depends upon both $x$ and $y$.  Furthermore,  when $z$ approaches the left or right side $C(z)$ goes to zero, so the factorization breaks down.

We have also simulated $C(z)$ with periodic boundary conditions on the horizontal sides.  Here, of course, all $P$'s are trivially independent of $y$ and therefore $C(z)$ is again a function of $x$ only (for either open or wired b.c.\ on the vertical sides).  $C(z)$ goes to a constant value different from $C_0$, viz. $C_1 \approx 1.022 \dots$ at points far from the ends of the cylinder.  We do not have a theoretical prediction for this value.

 Section \ref{nos} gives our numerical results which show approximate factorization and interesting $y-$dependence with wired b.c.\ , and section \ref{theory} presents our theoretical derivation of those results; first for the case of the semi-infinite strip (including open b.c.\ results and complete expressions for the densities) and then for the complete rectangle. We compare these predictions with further numerical results. 
 In section \ref{FScorrs} we consider the problem of finite-size corrections around an anchoring point.  We show that our formulas for wired b.c.\ predict these corrections.  In section \ref{periodicsec} we present our numerical results for periodic b.c., and section \ref{cons} gives our conclusions. The  appendix presents a full derivation of the semi-infinite strip densities, which confirms that our expressions for $P_L(z)$, $P_R(z)$, and $P_{L R}(z)$ are exact, as is  the $y-$independence of $C(z)$  with wired b.c.\,.

\section{Numerical results for C(z)} \label{nos}
To investigate the probabilities $P_L(z)$, etc., 
we carried out simulations in rectangular systems of dimensions $63 \times 127$, and $127 \times 255$.  (The arrays used in the computer code were actually exact powers of two, but one column and row were left open, along with periodic b.c., to efficiently simulate the open boundaries.)  We used a  square lattice and considered both bond and site percolation at the critical thresholds $1/2$ and $0.5927460$ 
\cite{NewmanZiff01,Lee08,FengDengBlote08} respectively.  The random-number generator used was R(471,1586,6988,9689) given in \cite{Ziff98}.  We kept track of the average density of clusters that touched the left, right, and both sides of the rectangular systems, where the density at a point is simply the number of times a cluster touching the desired boundary or boundaries includes that point, divided by the total number of trials.  In Fig.\ \ref{wiredfig} we show plots of the densities for bond percolation on a lattice of size $127 \times 255$ sites with wired boundary conditions on the left and right-hand sides.  The top figure shows $P_L(z)$; the plot of $P_R(z)$ is identical but flipped horizontally.  Along the left and right-hand boundaries $P_L(z)$ is constant and equal to $1$ and $\Pi_h$ respectively, both a consequence of the wired b.c.  In the intermediate region there is an exponential drop-off in the density.

In the center plot in Fig.\ \ref{wiredfig} we show $P_{L R}(z)$, which is roughly independent of $x$ away from the ends.  The lower figure shows $C(z)$ defined by Eq.\ (\ref{Cratio}), and here one can see the striking result that $C(z)$ depends upon the $x$-coordinate but appears to be independent of $y$, in spite of the strong $y$-dependence of the functions that define it.   Note that the range of the vertical scale now goes from $1$ to $1.03$.  

At the two wired boundaries $x = 0$ and $x = w$, $C(z)$ goes to 1.  For bond percolation, where all sites are effectively occupied while the bonds are diluted (occupied with probability $p$), $C(z)$ is identically 1 at these two boundaries because  $P_L(z) \to 1$, $P_{L R}(z) \to \Pi_h$, and $P_R(z) \to \Pi_h$ as $x \to 0$ for a given $y$ (and similarly for $x \to w$).  For site percolation, where sites including those in the first and last columns are occupied with probability $p = 0.592746\ldots$, we have $P_L(z) \to p$, $P_{L R}(z) \to p \Pi_h$, and $P_R(z) \to p \Pi_h$ as $x \to 0$ for a given $y$, so here $C(z) \to 1$ but only on the average, not identically as in the bond case, and there are small fluctuations.   We do not show the plots for site percolation as they are quite similar to those for bond percolation.  

Away from the left- and right-hand boundaries, $C(z)$ approaches the value $C_0$ given in Eq.\ (\ref{C0value}).  The quantity $C_0$ first appeared the context of the densities of clusters touching one or two boundary anchor points, where after just several lattice spacings away from the anchors the analogous $C(z)$ was found to go to $C_0$ everywhere \cite{KlebanSimmonsZiff06}.  The reason that the same constant appears here is that, from a distance, the interval looks like a point.  Furthermore, as discussed in more detail below, a conformal transformation converts the interval problem to the  point anchor problem, and shows that indeed $C(z)$ far from the vertical boundaries of the rectangle asymptotes to $C_0$.

\begin{figure}[htbp]
\begin{center}
\includegraphics[height=3in]{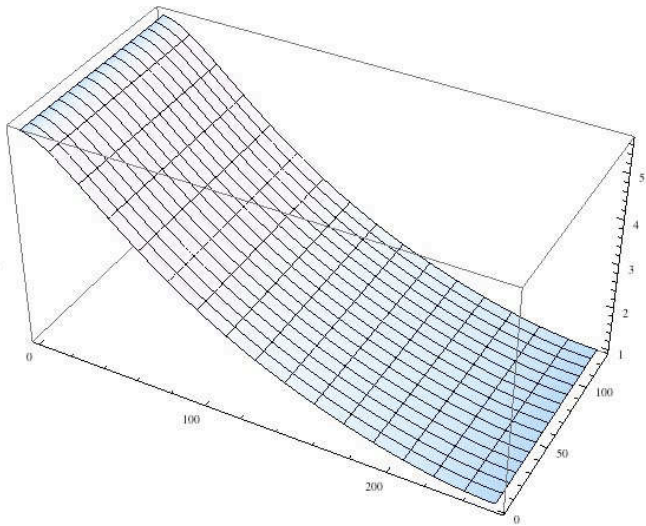}
\caption{Plot of $P_L(z)/P_{L R}(z)$ for the wired $127 \times 255$ system of Fig.\ \ref{wiredfig}.}
\label{PLoverPLR}
\end{center}
\end{figure}

Fig.\ \ref{PLoverPLR} shows a plot of the ratio $P_L(z)/P_{LR}(z)$ for the $127 \times 255$ system as shown in Fig.\ \ref{wiredfig}, showing quite clearly that this ratio is only a function of $x$ and not $y$ for wired b.c.  Likewise, $P_L(z)/P_R(z)$  only depends on $y$  (this follows because $C(z)$ is only a function of $y$).  Thus, all three quantities $P_L(z)$, $P_R(z)$, and $P_{LR}(z)$ have a common $y$ dependence for a given value of $x$.  We will use this to motivate our theoretical results.

\begin{figure}[htbp]
\begin{center}
\includegraphics[height=5in]{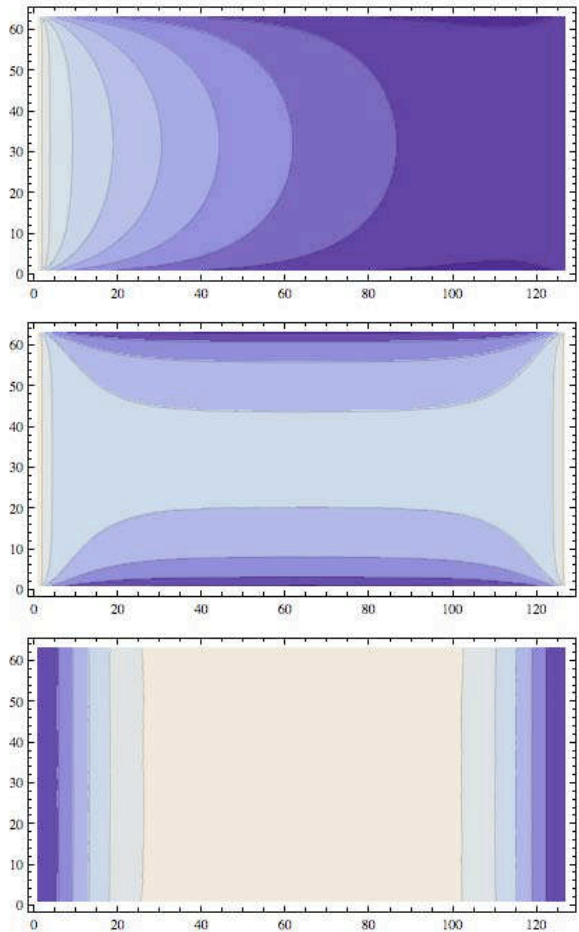}
\caption{Contour plots of $P_L(z)$ (top), $P_{L R}(z)$ (center), and $C(z)$ (bottom) for a $63 \times 127$ system
with wired b.c.\ on the left and right-hand sides (similar to Fig.\ \ref{wiredfig} but here as contour plots).  In the plot of $C(z)$ (lower figure), the contours are at $1.005$, $1.010$, $1.015$, $1.020$, and $1.025$.  The lower figure may be compared with Fig.\ \ref{opencontours}, which shows a similar plot for open b.c.\, where the behavior is quite different.}
\label{contourplot}
\end{center}
\end{figure}

In Fig.\ \ref{contourplot} we show similar data for a $63 \times 127$ bond percolation system, this time plotted as contour plots.  The apparent $y$ independence of $C(z)$ is quite clear. For much smaller system sizes  (e.g.\ $15 \times 31$), finite-size effects do lead to visible curvature in the contours of $C(z)$ (not shown here).

\begin{figure}[htbp]
\begin{center}
\includegraphics[scale=0.4]{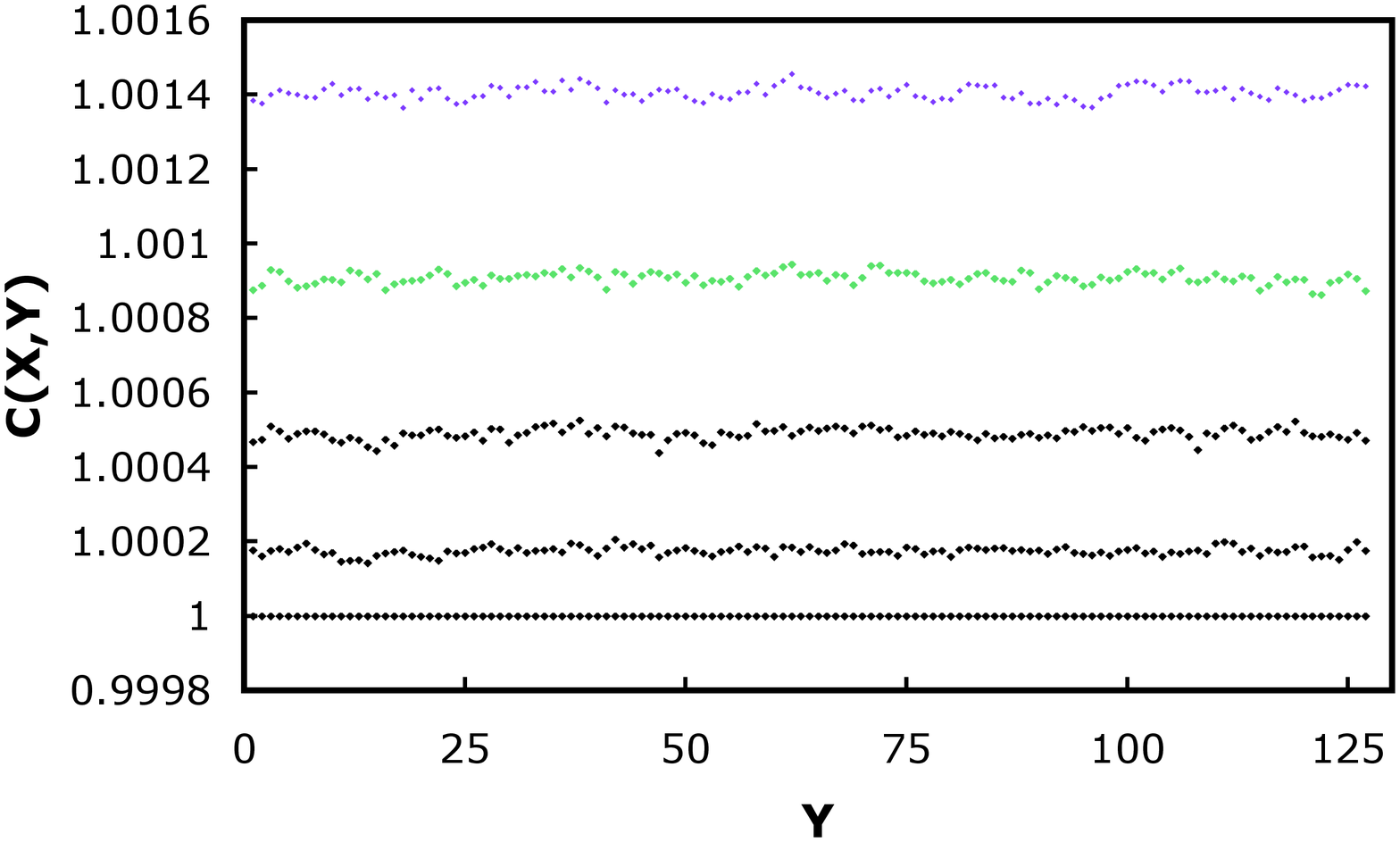}
\includegraphics[scale=0.4]{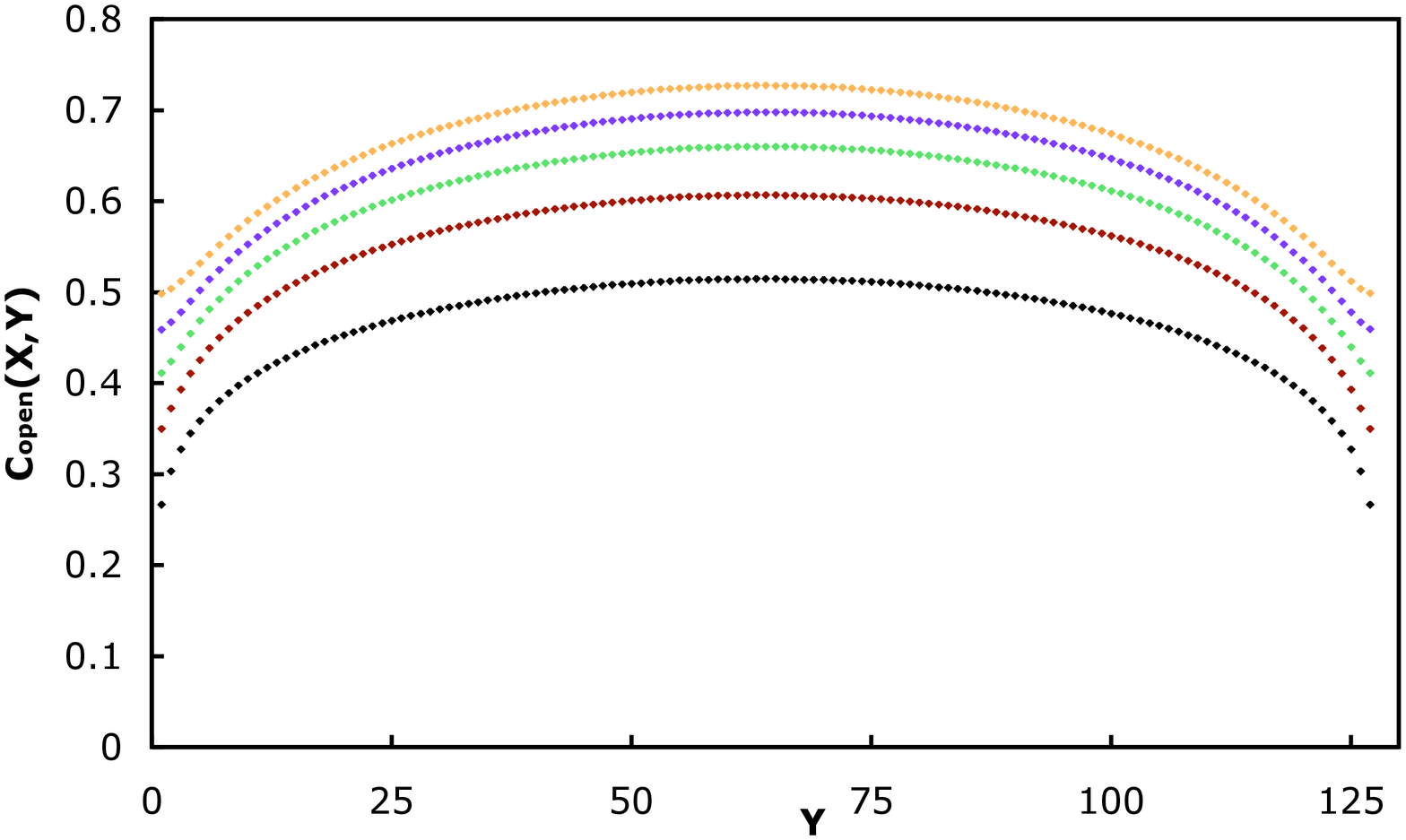}
\caption{Top figure: $C(z)$ as a function of $Y$ for $X = 1, 2, 3, 4, 5$ (bottom to top) for a system of size height$\times$width = $127 \times 255$, 
with wired b.c.\ on the sides. Bottom figure: same simulation with open b.c.\  (Note the change of vertical scale.)}
\label{edgefig}
\end{center}
\end{figure}

The data for $x = 1,\ldots, 5$ for a system of size $127 \times 255$ are shown in the upper plot in Fig.\ \ref{edgefig}, showing the fluctuation in the data about the constant values.  These rather large fluctuations are due to the fact that the probabilities $P_R(z)$ and $P_{L R}(z)$ are quite small for $z$ near the sides.   Here $1.14 \cdot 10^9$ samples were generated, a similar number as for the other plots.  The lower plot shows the same simulation with open b.c.\ for comparison; the dependence on b.c.\ is readily apparent.

\begin{figure}[htbp]
\begin{center}
\includegraphics[scale=0.4]{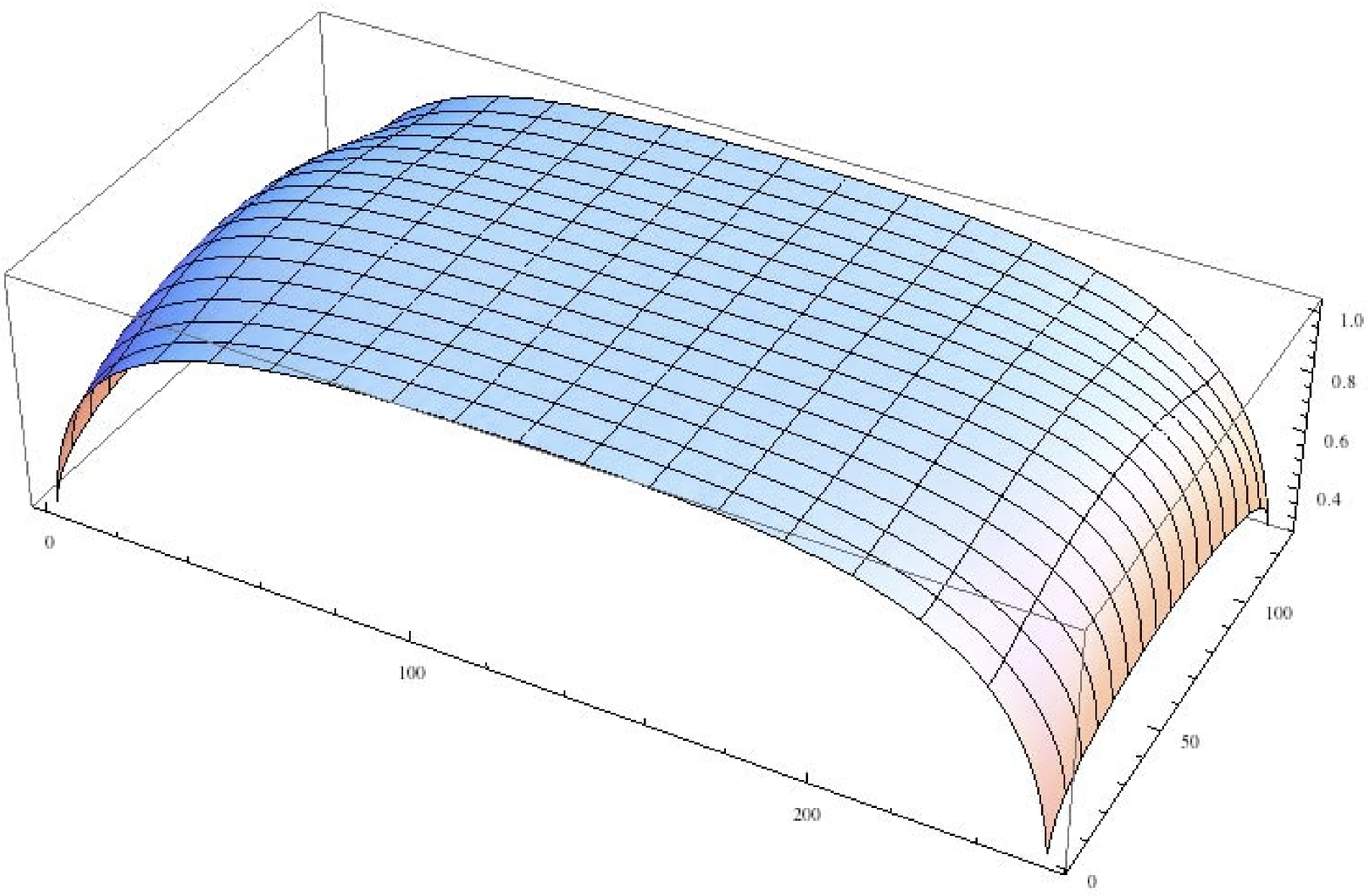}
\caption{Plot of numerical results for $C_{\mathrm{open}}(z)$ for a system of size $127 \times 255$ with open boundaries on all sides.  In contrast to the wired case, Fig\ \ref{wiredfig}, $C_{\mathrm{open}}(z)$ goes to a value much less than 1 at the ends, going to zero as the system size increases to infinity.  The dependence of $C_{\mathrm{open}}(z)$ upon the $y$ coordinate is evident here. }
\label{Copen}
\end{center}
\end{figure}

In Fig.\ \ref{Copen} we show $C_{\mathrm{open}}(z)$ for the case of open boundary conditions on the left and right-hand sides. Here the behavior differs markedly.  While $C_{\mathrm{open}}(z)$ still approaches the value $C_0$ away from the sides, closer to them a significant $y$-dependence of $C(z)$ can be seen.  Also, $C(z)$ does not go to $1$ at those sides as it does in the wired case, but rather drops to $0$, highlighting the sensitivity of the factorization to boundary conditions. 

\section{Theoretical results} \label{theory}
In this section we use boundary conformal field theory to calculate  $C(z)$.  This quantity (see (\ref{Cratio}))  is given by a ratio of various densities.  Our work in \cite{KlebanSimmonsZiff06,SimmonsKlebanZiffPRE07} illustrates the method by which such densities can be derived from correlation functions in  conformal field theory with $c=0$.   The three densities we are interested in arise as different conformal blocks for the same set of operators, being distinguished by fusion rules, as we will see.   Thus in the upper half-plane $\mathbb{H}$, with $w=u+i v$,
\be \label{PLRz}
P_{L R}(w), P_L(w), P_R( w)   \sim \langle\phi_{1,2}(u_1) \phi_{1,2}(u_2) \psi(w_3,\bar w_3) \phi_{1,2}(u_4) \phi_{1,2}(u_5) \rangle \; ,
\ee
where the boundary operator $\phi_{1,2}(u)$, with conformal dimension $h_{1,2}=0$, implements a change from open to wired (free to fixed) boundary conditions at the point $u$, and  the ``magnetization" operator $\psi(w) := \phi_{3/2,3/2}(w)$, with conformal dimension $h_{\psi}=\bar h_{\psi}=5/96$, measures the density of clusters at a point $w$ in the upper half-plane.   The other quantity of interest, $\Pi_h$,  is simply the probability that there exists a crossing cluster in the rectangle, given by Cardy's formula (which itself follows from a correlation function like (\ref{PLRz}) but without the $\psi$ operator) \cite{Cardy92}.

These correlation functions can be constructed from the conformal blocks allowed by (\ref{PLRz}). However, computing these conformal blocks is a formidable task involving a six-point correlation function that depends on three independent cross-ratios.  We can make the calculation tractable  by exploiting the observed independence  on the vertical coordinate in the rectangle: $C(x,y)=C(x)$.  Letting $y_3 \to 0$ (so $v_3 \to 0$ in $\mathbb{H}$),  the bulk magnetization operator $\psi$ in $\mathbb{H}$ is replaced with a boundary magnetization operator, $\phi_{1,3}(u_3)$ so that
\be \label{PLRz2}
P_{L R}(u), P_L(u), P_R(u)    \sim \langle\phi_{1,2}(u_1) \phi_{1,2}(u_2) \phi_{1,3}(u_3) \phi_{1,2}(u_4) \phi_{1,2}(u_5) \rangle \; .
\ee
Note that (\ref{PLRz2}) is a five-point function involving only boundary operators that we will compute shortly. In this section, we assume $y$-independence throughout.  However, the appendix shows that our results are valid without this assumption in the case of the semi-infinite strip.

We next determine  (\ref{PLRz2})  in the upper half-plane, and then transform the result into the appropriate geometry. (This transformation leaves $C$ invariant \cite{KlebanSimmonsZiff06}.)

In subsections \ref{SIStrip} and \ref{FinRect} we find closed expressions for $P_{L R}(x)$, $P_L(x)$, and $P_R(x)$ in the semi-infinite strip and rectangle.  The appendix verifies that for the case of a semi-infinite strip these expressions are valid without assuming $y$-independence.

Note that the variables  $z$ and $x$ represent  bulk and boundary densities in the rectangle or strip while $w$ and $u$ represent bulk and boundary densities in $\mathbb{H}$. 

\subsection{Semi-infinite strip} \label{SIStrip}

We next consider the  semi-infinite strip, which approximates the behavior near the end of a long rectangle. This is equivalent to replacing the right-hand interval with a point operator because it is vanishingly likely that two distinct clusters emanate from the distant end.   Depending on which of the densities we are calculating we make use of the fusion rules to replace $\phi_{1,2}(u_4) \phi_{1,2}(u_5)$ with either $\mathbf{1}(\infty)$ or $\phi_{1,3}(\infty)$, which significantly simplifies the calculation.

Because $\phi_{1,3}$ is the boundary magnetization operator,  the fusion $\phi_{1,2} \times \phi_{1,2} = \phi_{1,3}$ conditions the wired interval to connect to other objects marked in the correlation function.  For example, to calculate $P_R(u)$ and $P_{L R}(u)$ and then transform the result into the semi-infinite strip we insert the $\phi_{1,3}(\infty)$ operator, because these two quantities require that a cluster connect the point at infinity with $u$.  (Here and in the remainder of this section we ignore the scaling factor and operator product expansion (OPE) coefficient in this fusion since they cancel in the ratios that we calculate.)

The fusion $\phi_{1,2} \times \phi_{1,2} = \mathbf{1}$ places no condition on the connectivity of the wired interval.  So we get configurations with the interval isolated and with it connected to other points.   Note that the identity fusion encompasses all of the configurations in the $\phi_{1,3}$ bulk-boundary fusion as well.  This mirrors Cardy's derivation of the horizontal crossing probability, where $\Pi_h$ corresponds to the $\phi_{1,3}$ conformal block while the identity block  counts all configurations including those with a horizontal crossing \cite{Cardy92}.

As mentioned, one can calculate $P_R(u)$ and $P_{L R}(u)$ using the correlation function
\be\label{CF}
\langle \phi_{1,2}(u_1) \phi_{1,2}(u_2) \phi_{1,3}(u_3) \phi_{1,3}(u_4) \rangle = (u_4-u_3)^{-2/3} F\left( \frac{(u_2-u_1)(u_4-u_3)}{(u_3-u_1)(u_4-u_2)}\right)\; .
\ee 
The null-state for $\phi_{1,2}$ gives
\be\label{DE}
0=\left(\frac{1/3}{(u_4-u_1)^2}+\frac{1/3}{(u_3-u_1)^2}-\frac{\partial_{u_4}}{u_4-u_1}-\frac{\partial_{u_3}}{u_3-u_1}-\frac{\partial_{u_2}}{u_2-u_1} -\frac{3}{2}\partial_{u_1}^2\right)\langle \phi_{1,2}(u_1) \phi_{1,2}(u_2) \phi_{1,3}(u_3) \phi_{1,3}(u_4) \rangle\; .
 \ee
Applying (\ref{DE}) to (\ref{CF}) and using conformal symmetry to take $\{ u_1,u_2,u_3,u_4\}$ $\Rightarrow$ $\{0, \l, 1,  \infty\}$ in the usual manner, so that $\l$ is the cross-ratio, we find 
\be
0=F''(\l)+\frac{2(1-2\l)}{3\l(1-\l)}F'(\l)-\frac{2}{9(1-\l)^2}F(\l)\; .
\ee
The solutions  yield the conformal blocks
\bea \label{CB1}
F_{\mathbf{1}}(\l)&=&(1-\l)^{2/3}{}_2F_1(1/3, 4/3, 2/3 , \l )\; ,\quad \mathrm{and}\\ 
F_{\phi_{1,3}}(\l)&=&\l^{1/3}(1-\l)^{2/3}{}_2F_1( 2/3, 5/3, 4/3 , \l )\; . \label{CB2}
\eea
Here the index on the blocks represents the corresponding fusion channel of the $\phi_{1,2}$ operators.

\begin{figure}[htbp]
\begin{center}
\includegraphics[scale=0.6]{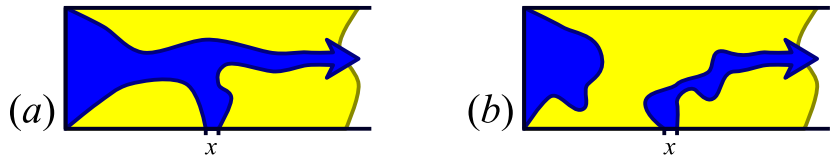}
\caption{Schematic representation of the two types of cluster configurations consistent with $\langle \phi_{1,2}(0) \phi_{1,2}(\l) \phi_{1,3}(1) \phi_{1,3}(\infty) \rangle$,  translated into the strip geometry.}
\label{SIstripconfig}
\end{center}
\end{figure}

\begin{figure}[htbp]
\begin{center}
\includegraphics[scale=0.6]{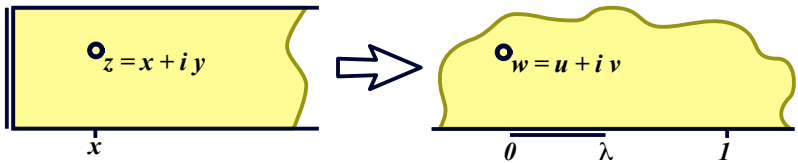}
\caption{Semi-infinite strip mapping.}
\label{KeSIStrip}
\end{center}
\end{figure}

By the above, we may identify the  configurations with $(0, \l)$ connected to $1$ and infinity with the conformal block $F_{\phi_{1,3}}(\l)$.  Upon mapping into the strip this correlation will become $P_{L R}(x)$ as illustrated in Fig.\ \ref{SIstripconfig}(\emph{a}).

The conformal block $F_{\mathbf{1}}(\l)$ represents the  configurations with $(0,\l)$ isolated from, as well as connected to, $1$ and infinity.  Thus upon mapping to the strip (see Fig.\ \ref{KeSIStrip}) this block becomes $P_R(x)$, which includes all configurations with $x$ connected to $R$, as illustrated in Fig.\ \ref{SIstripconfig}(\emph{a}\&\emph{b}).  Note that $F_{\mathbf{1}}(\l)$ (in $\mathbb{H}$) includes configurations where $u$ and $R$ are not directly connected, but instead connect through the wired boundary conditions on the left side.  This subtlety will be discussed in more detail in section \ref{SIStripDens}.

Having found $P_{L R}(u)$ and $P_{R}(u)$ we need only determine $P_L(u)$.  Configurations of this type don't need to cross to the right-hand side, so instead of placing $\phi_{1,3}$ at infinity we place an identity; thus $P_L(u)$ is proportional to a three point function.

Thus, including the proper normalizations and scaling factors we find
\bea \label{Prob1}
P_{L R}(u)&=&\langle \phi_{1,2}( 0 ) \phi_{1,2}(\l)_{[1,3]}\phi_{1,3}(1) \phi_{1,3}(\xi)\rangle \quad=\quad C_{1 1 2} C_{2 2 2} \xi^{-2/3} F_{\phi_{1,3}}(\l)\; ,\\ \label{Prob2}
P_L(u)&=& \langle \phi_{1,2}(0) \phi_{1,2}(\l)  \phi_{1,3}(1)\rangle \phantom{\phi_{1,3}(\xi)_{[1,2]}} \quad=\quad C_{1 1 2}\l^{1/3}(1-\l)^{-1/3}\; ,\\ \label{Prob3}
\Pi_h&=& \langle \phi_{1,2}( 0 ) \phi_{1,2}(\l) \phi_{1,3}(\xi)\rangle \phantom{\phi_{1,3}(1)_{[1,2]}} \quad=\quad C_{1 1 2}\l^{1/3}\xi^{-2/3}\; \qquad\mathrm{and},\\
P_R(u)&=&\langle \phi_{1,2}( 0 )\phi_{1,2}(\l)_{[1,1]}  \phi_{1,3}(1) \phi_{1,3}(\xi)\rangle \quad=\quad \xi^{-2/3} F_{\mathbf{1}}(\l)\; ,  \label{Prob4}
\eea
where the variable $\xi (\to \infty)$ indicates the order of magnitude of the length of the rectangle, the bracketed subscript notation denotes the fusion channel propagating between the operators separated (following Bauer and Bernard \cite{BauerBernard03}), and the indices on the (boundary) operator product expansion coefficients $C_{i j k}$ follow the convention of \cite{SimmonsKleban07} with index $j$ representing the $j-$leg boundary operator $\phi_{1,1+j}$.  Both operator product expansion coefficients that appear above are previously known: $C_{1 1 2}{}^2 = 2 \pi \sqrt{3} / \Gamma(1/3)^3 = 0.56604668\ldots$ is most familiar as the leading coefficient in Cardy's formula \cite{Cardy92}, while $C_{2 2 2}=C_0$ arose in the factorization formula in \cite{KlebanSimmonsZiff06,SimmonsKlebanZiffPRE07}.  Note that the expressions for $P_L(u)$ and $\Pi_h$  could also be obtained as limiting cases of Cardy's formula when one interval shrinks to a point.

For future reference, note that if the interval $L$ is not connected to either $u$ or $R$ we have 
\be \label{preXsym}
P_{\bar L R}(u) = \langle \phi_{1,3}(1) \phi_{1,2}(\l)_{\partial [1,2]}\phi_{1,2}( 0 ) \phi_{1,3}(\xi)\rangle= K_{\partial \phi_{1,2}} \xi^{-2/3} G_{\partial \phi_{1,2}}(\l)\; ,
\ee
where
\be
G_{\partial \phi_{1,2}}(\l)= (1-\l)^{2/3}{}_2F_1(1/3, 4/3, 2 , 1-\l )\qquad \mathrm{and}\qquad K_{\partial \phi_{1,2}} = \frac{8\, \pi^2}{9\, \Gamma(1/3)^3}\; .
\ee
This can be deduced by observing that $P_{\bar L R}(u)=P_R(u)-P_{L R}(u)$ and using the crossing symmetry relation for this correlation function,
\be \label{Xsym}
K _{\partial \phi_{1,2}}G_{\partial \phi_{1,2}}(\l) = F_{\mathbf{1}}(\l)- C_{1 1 2} C_{2 2 2} F_{\phi_{1,3}}(\l)\; .
\ee
In (\ref{preXsym}) we label the propagation channel and conformal block with $\partial \phi_{1,2}$ in order to be consistent with the underlying logarithmic conformal field theory  \cite{SimmonsKlebanZiffJPA07}. However, for present purposes the fusion $\phi_{1,3} \times \phi_{1,2}$ is more complicated to interpret physically than the $\phi_{1,2} \times \phi_{1,2}$ fusion discussed above, so while we include the notation for accuracy, the distinction is unimportant here and in what follows.

Combining (\ref{Cratio}) and (\ref{CB1})--(\ref{Prob4}) we find
\be\label{ApproxUHP}
C(\l) = C_0 \sqrt{1-\l} \;\frac{ {}_2F_1(2/3,5/3;4/3;\l)}{\sqrt{{}_2F_1(1/3,4/3;2/3;\l)}}\; .
\ee

Now map from the semi-infinite strip $\mathbb{S}=\{ z=x+i y\,|\, 0<x,\;0<y<1\}$ via 
\be \label{sismap}
w(z) = \frac{\cosh^2(\pi z/2)}{\cosh^2(\pi x/2)}\; ,
\ee
which takes the point $x$ on the lower side of the strip to $1$ in the upper half-plane (see Fig.\ \ref{KeSIStrip}).

The cross-ratio is given by
\be
\l=w(0)=\mathrm{sech}^2(\pi x/2)\; .
\ee
Using (\ref{ApproxUHP}) and standard hypergeometric identities involving quadratic arguments \cite{AbSt} gives
\be \label{ApproxSIStrip}
C(x) = C_0\frac{{}_2F_1(  -1/2,-1/3,7/6, e^{-2 \pi x} )}{\sqrt{{}_2F_1( -1/2,-2/3,5/6  , e^{-2 \pi x})}} \; .
\ee
The appendix shows that (\ref{ApproxSIStrip}) is in fact exact, and follows without the assumption of $y$-independence.

Now far from the finite end of the strip $C(x)$ must reduce to the analogous result involving the density of percolation clusters constrained to touch two isolated boundary points \cite{KlebanSimmonsZiff06}.  Thus  $C_0 = C_{2 2 2}$ and (\ref{C0value}) follow.   Alternately, if $x=0$ then (as mentioned) the wired boundary conditions ensure that $P_L(z)=1$ and $P_{L R}(z) = \Pi_h=P_R(z)$ so that  $C(0)=1$.  
It is a non-trivial check of (\ref{ApproxSIStrip}) that these limiting cases are indeed consistently recovered.

\begin{figure}
\begin{center}
\includegraphics[width=5in]{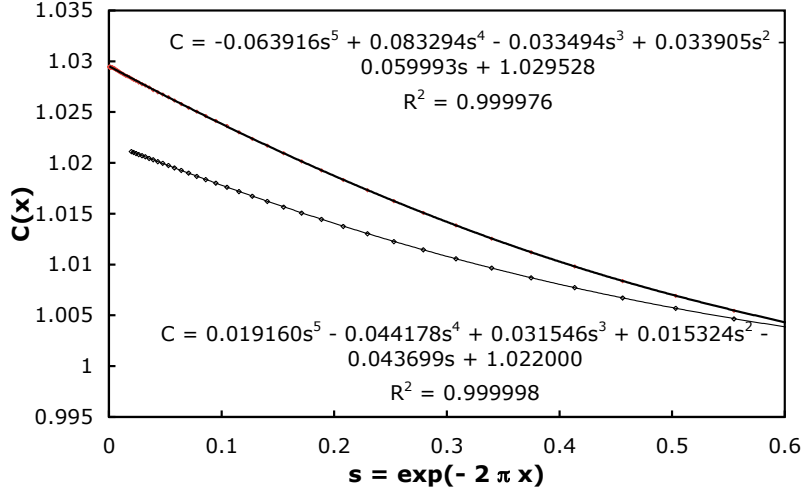}
\caption{$C(x)$ vs.\ $s = e^{-2 \pi x}$ for wired b.c.\ on the vertical sides, for open b.c.\ (upper curve) and periodic (lower curve) b.c.\ on the horizontal sides, for systems of size $63 \times 255$ (open) and $64 \times 128$ (periodic). Here $x = (X-1)/63$ (open) and $(X-1)/64$ (periodic), where $X$ is the lattice coordinate of the column.  Data are fit to fifth-order polynomials as shown.  For the open case, the first few terms compare favorably with the prediction, Eq.\ (\ref{ApproxSIStripAsymp}).  For the periodic case, we have no theoretical prediction.}
\label{CvsExpBoth}
\end{center}
\end{figure}

Expanding (\ref{ApproxSIStrip}), we find
\be \label{ApproxSIStripAsymp}
C(x)=C_0\left(1-\frac{2}{35} e^{-2 \pi x}+\frac{834}{25025} e^{-4 \pi x}-\frac{6406}{734825} e^{-6 \pi x}+\mathcal{O}(e^{-8 \pi x})\; ,\right)\; .
\ee
In Fig.\ \ref{CvsExpBoth} (upper curve), we compare this prediction with simulation results from a system of size $63 \times 255$.  Fitting the numerical results to a polynomial in $s = \exp(-2 \pi x)$, we find $C(x) = 1.02953 - 0.059993 s + 0.033494 s^2\ldots = 
1.02953(1 - 0.058272 s + 0.032933 s^2 \ldots)$ while (\ref{ApproxSIStripAsymp}) gives $C_0(1 - 0.0571429 s + 0.0333267 s^2 \ldots$), so the agreement is quite good.

\subsection{Density in the semi-infinite strip}\label{SIStripDens}

In the previous section, based on numerical evidence, we assumed that the ratio $C(z)$ in the semi-infinite strip was independent of vertical position and performed the calculation in the simpler case when $z$ is on the boundary.  However, the simulations further suggest that any ratio of two of $P_{L R}(z)$, $P_L(z)$, and $P_R(z)$ is also independent of $y$, as shown in Fig.\ref{PLoverPLR}.  Given this stronger condition (which, for the semi-infinite strip, is proven in the appendix), knowing any one of these three functions immediately determines the other two via the expressions in section \ref{SIStrip}. In this section we exploit this idea to find $P_{L R}(z)$, $P_L(z)$, and $P_R(z)$ and also the corresponding quantities for open b.c.

In the semi-infinite strip we can easily calculate $P_L(z)$ based on results for the density of clusters anchored to an interval in \cite{KlebanSimmonsZiff06}.  As the side $R$ is taken to $\infty$, the probability of spanning the length of the strip is negligible compared to the probability of connecting to side $L$.  Thus there is no distinction between wired and open boundary conditions on $R$, and we can equate $P_L(z)$, with $p_L(z)$, the density of clusters attached to the left side when the sides have open boundary conditions. (In what follows we differentiate between the boundary conditions for our densities by using  `$p$' to refer to open sides and  `$P$' to refer to wired sides.)

The density of clusters at a point $w=u+i v$ in the upper half-plane that are attached to an interval $I=(u_1,u_2)$ is  \cite{KlebanSimmonsZiff06}
\be
p_I(w) \sim v^{-5/48}\left(\frac{2-\eta}{2\sqrt{1-\eta}}-1\right)^{1/6}\; ; \quad \eta=\frac{(u_2-u_1)(\bar w-w)}{(w-u_1)(\bar w -u_2)}\; .
\ee
We next transform this into the semi-infinite strip using the mapping (\ref{sismap}).   The points $\{i,0,z\}$ in the strip map to the points $\left\{0,\l,\l \cosh^2\left(\frac{\pi}{2} (u+i v) \right) \right\}$ in the upper half-plane so that the corresponding cross-ratio is
\be
\eta 
= \frac{(\l-0)\left(\l \cosh^2\left(\frac{\pi}{2} (x-i y) \right) -\l \cosh^2\left(\frac{\pi}{2} (x+i y) \right)\right)}{\left(\l \cosh^2\left(\frac{\pi}{2} (x+i y) \right)-0\right)\left(\l \cosh^2\left(\frac{\pi}{2} (x-i y) \right) -\l\right)}
=1- \left( \frac{\sinh(\pi x)+i  \sin(\pi y)}{\sinh(\pi x)-i  \sin(\pi y)}\right)^2\; .
\ee
This leads to 
\be \label{PLw}
P_L(z)=p_L(z)\sim\frac{1}{\sinh^{5/48}(\pi x)} \left( \frac{\sin^2(\pi y)}{\pi^2(\sinh^2(\pi x)+\sin^2(\pi y))}\right)^{11/96}\; .
\ee

The expression for $P_L(z)$ then determines the other two densities.
Using  (\ref{Prob1}),  (\ref{Prob2}), and  (\ref{Prob4}) and including all relevant transformation factors we find, for points $x$ on the side of the strip,
\bea \label{ProbSI11}
P_{L R}(x)&=& \Pi_h C_{2 2 2} \left( \frac{\pi}{1-e^{-2 \pi x}} \right)^{1/3}{}_2F_1\left(-\frac{1}{2}, -\frac{1}{3}, \frac{7}{6} , e^{-2 \pi x} \right)\; ,\\
P_L(x)&=&C_{1 1 2}\left( \frac{4 \pi e^{-\pi x}}{1-e^{-2 \pi x}}\right)^{1/3}\;, \quad \mathrm{and}\\
P_R(x)&=&\frac{\Pi_h}{C_{1 1 2}}\left( \frac{\pi}{4e^{-\pi x}(1-e^{-2 \pi x})} \right)^{1/3}{}_2F_1\left(-\frac{2}{3}, -\frac{1}{2}, \frac{5}{6} , e^{-2 \pi x} \right)\; ,  \label{ProbSI13}
\eea
where  $\l = \mathrm{sech}^2( \pi x/2)$, $w'(x) = \pi \sqrt{1-\l}$, and  hypergeometric identities have been used.  
Thus
\bea \label{PLwR}
P_{L R}(z)\;  
&=&\frac{\Pi_h P_L(z)\, C_{2 2 2}\,e^{\pi x/3}}{C_{1 1 2}\; 4^{1/3}}{}_2F_1\left(-\frac{1}{2}, -\frac{1}{3}, \frac{7}{6} , e^{-2 \pi x} \right)\; ,\; \mathrm{and}\\
P_R(z)\; 
&=&\frac{\Pi_h P_L(z)\,e^{2 \pi x/3}}{C_{1 1 2}{}^2\; 4^{2/3}}{}_2F_1\left(-\frac{2}{3}, -\frac{1}{2}, \frac{5}{6} , e^{-2 \pi x} \right)\; , \label{PwR}
\eea
where $P_L(z)$, given in (\ref{PLw}), contains all the $y$ dependence in these expressions, and $\Pi_h$ is given in (\ref{Prob3}).

These densities apply with wired boundary conditions on the left side, but the analogous expressions can be found for open boundary conditions.   One more result is required: $p_R(z)$, the density of clusters attached to the distant right-hand side when the left side has open boundary conditions.
In the upper half plane this quantity is given by
\be
p_R(w)=\langle \psi(w,\bar w) \phi_{1,3}(\xi) \rangle = v^{-5/48}\left( \frac{v}{2 |w-\xi|^2}\right)^{1/3}\; ,
\ee
with $\xi \to \infty$, which becomes
\bea
p_R(z)&=&\left(\frac{|w'(z)|}{v}\right)^{5/48}w'(\g)^{1/3} \xi^{-2/3} (v/2)^{1/3}\\
&=&\frac{\Pi_h}{(4 \pi)^{1/3}C_{1 1 2}{}^2} \left( \pi^2(\sinh^2(\pi x)+\sin^2(\pi y))\right)^{5/96} \left( \sinh(\pi x) \sin(\pi y) \right)^{11/48}\\
&=&\frac{\Pi_h P_L(z)}{C_{1 1 2}{}^2\;4^{1/3}} \sinh^{1/3}(\pi x)\left(\sinh^2(\pi x)+\sin^2(\pi y)\right)^{1/6} \; , \label{pwR}
\eea
when transformed into the strip. The last equation is included to simplify comparison with (\ref{PwR}).  The normalization of $p_R(x)$ is consistent with that chosen for (\ref{PLw}) (the same non-universal factors are omitted).  A check of this is that $P_R(z)-p_R(z) \to 0$ as $x \to \infty$, which we expect since the boundary conditions on the left hand side are unimportant in that limit.

With this result we can calculate the density of crossing clusters, $p_{L R}(z)$ with open boundary conditions on the left hand side.  Note that the density of clusters that attach to the distant right side but not the left is independent of the particular boundary conditions we place on the left side, and as discussed the boundary conditions on the right-hand side are unimportant in the strip.  This implies that $P_{\bar L R}(z)  = p_{\bar L R}(z)$, with $P_{\bar L R}(z)=P_R(z)-P_{L R}(z)$ and $p_{\bar L R}(z)=p_R(z)-p_{L R}(z)$, where, e.g.\  $P_{\bar L R}(z)$ is the density at $z$ of clusters that connect to $R$ but not to $L$.
An explicit result for this quantity   
\be
p_{\bar L R}(z) =P_{\bar L R}(z) =\frac{\Pi_h P_L(z)\, K_{\partial \phi_{1,2}} \,(1-e^{-2 \pi x})^2}{16\; e^{-2 \pi x/3}\; C_{1 1 2}{}^2}{}_2F_1\left(\frac{4}{3}, \frac{3}{2}, 3, 1-e^{-2 \pi x} \right)
\ee
follows from (\ref{PLwR}) and (\ref{PwR}), and leads to
\bea 
p_{L R}(z)
&=&\frac{\Pi_h P_L(z)\sinh^{1/3}(\pi x)}{C_{1 1 2}{}^2} \left(\frac{\left(\sinh^2(\pi x)+\sin^2(\pi y)\right)^{1/6}}{4^{1/3}}\right.\\
&&\phantom{\frac{\Pi_h P_L(z)\sinh^{1/3}(\pi x)}{C_{1 1 2}{}^2}}\left.-\frac{K_{\partial \phi_{1,2}}\; \tanh^{5/3}(\pi x)}{4}{}_2F_1\left(\frac{4}{3}, \frac{5}{6}, 2,\tanh^2(\pi x) \right) \right) \; , \nonumber
\eea
after simplification using hypergeometric identities.

\begin{figure}[htbp]
\begin{center}
\includegraphics[scale=0.4]{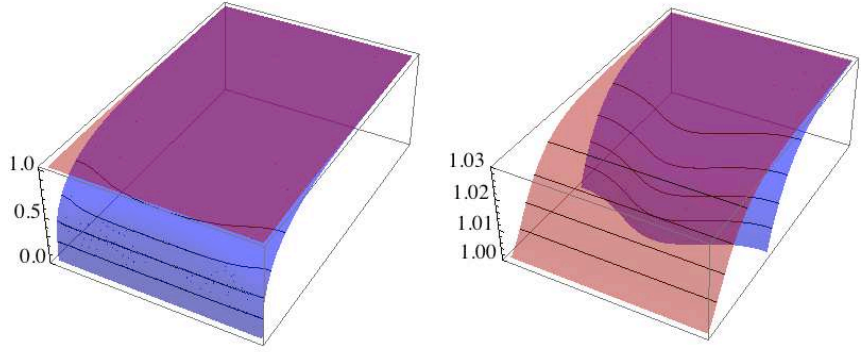}
\caption{Theoretical prediction for $C(z)$ (upper, red; Eq.\ (\ref{ApproxSIStrip})) and $C_{\mathrm{open}}(z)$ (lower, blue; Eq.\ (\ref {openstripeq})) near the end of an infinite strip, which are comparable
to the simulation results shown in Figs.\ \ref{wiredfig}, \ref{edgefig} and \ref{Copen}.  The two plots differ only in scale: the right-hand plot focuses on the range of $C(z)$ which is obscured on the left.}
\label{jakesopen}
\end{center}
\end{figure}

Thus for a strip with open boundary conditions we have
\bea \nonumber
C_{\mathrm{open}}(x,y)&=&\frac{p_{L R}(z)}{\sqrt{p_L(z)p_R(z) \Pi_h}}\\ \nonumber
&=&\frac{\sinh^{1/6}(\pi x)\left(\sinh^2(\pi x)+\sin^2(\pi y)\right)^{1/12}}{C_{1 1 2}\; 2^{1/3}}\\  
&&-\frac{K_{\partial \phi_{1,2}}\;\sinh^{1/6}(\pi x)\; \tanh^{5/3}(\pi x)}{C_{1 1 2}\; 2^{5/3}\; \left(\sinh^2(\pi x)+\sin^2(\pi y)\right)^{1/12}}{}_2F_1\left(\frac{4}{3}, \frac{5}{6}, 2, \tanh^2(\pi x)\right)\; .
\label{openstripeq}
\eea
This expression is considerably more complicated than (\ref{ApproxSIStrip}), and exhibits a dependence on $y$ that is not present with wired ends, as shown in Fig.\ \ref{jakesopen}.  In addition $0<C_{\mathrm{open}}(x,y)<C_{2 2 2}$ while $1<C(x)<C_{2 2 2}$  so that factorization is a much less accurate approximation near an open anchoring interval.

\begin{figure}[htbp]
\begin{center}
\includegraphics[scale=0.4]{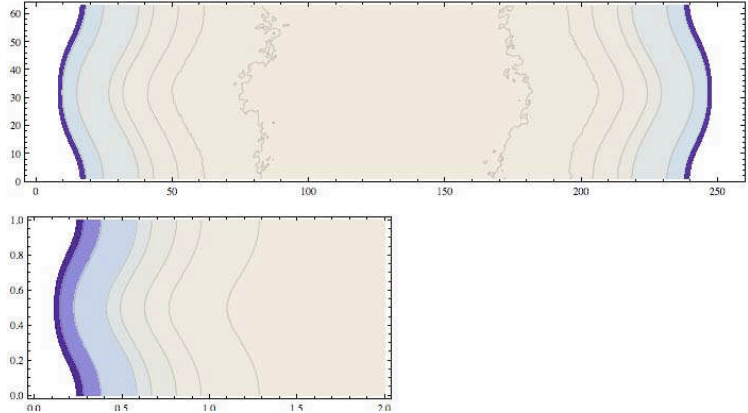}
\caption{Simulation results (above) of $C_\mathrm{open}(z)$ on a rectangle of size $63 \times 127$, compared with the theoretical prediction of (\ref{openstripeq}) for a strip.  Contours are at $C = 1.029, 1.025, 1.02, 1.01, 1.0, 0.95$ and $0.9$, going toward the ends.}
\label{opencontours}
\end{center}
\end{figure}
To compare this result with simulations, we considered a system of size $63 \times 255$ with open b.c.\ on all four sides, generating $10^8$ samples. 
In Fig.\ \ref{opencontours}, we show the contours predicted by Eq.\ (\ref{openstripeq}) (lower figure) and compare them with the contours of the numerical simulations of the $63 \times 255$ system (upper); clearly the behavior is similar.   This may also be compared with Fig.\ \ref{contourplot}, which shows the contour lines with wired b.c.

\begin{figure}[htbp]
\begin{center}
\includegraphics[scale=0.4]{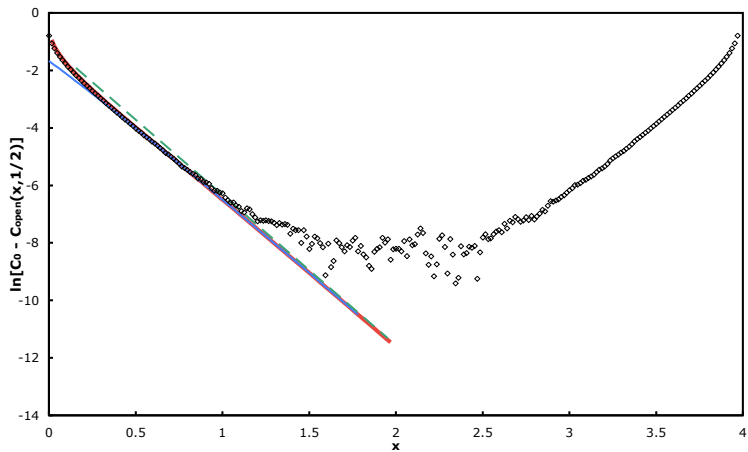}
\caption{Results of $\ln[C_0 - C_\mathrm{open}(x,1/2)]$ vs.\ $x$ along the centerline $y = 1/2$ for a lattice of size $63 \times 255$ (data points) compared with
theoretical prediction for the half-infinite strip, Eq.\ (\ref{openstripeq}) (red solid line), one term (blue dashed line) and
two terms (green dotted line) in the asymptotic expansion, Eq.\ (\ref{openasymp}). }
\label{bobsopen}
\end{center}
\end{figure}

To make a more quantitative comparision, in Fig.\ \ref{bobsopen} we compare the results at one end of the rectangular system with the theoretical strip results, Eq.\ (\ref{openstripeq}).  Asymptotically for large $x$, Eq.\ (\ref{openstripeq}) behaves as
\begin{equation}
C_{\mathrm{open}}(x,y) = C_0 \left( 1 - \frac{\sqrt{3}\Gamma\left(\frac{1}{3}\right)^6}{2^{2/3} \pi^3}
\frac{11 + 5 \cos(2 \pi y)}{240}e^{-5 \pi x / 3} +\frac{13 + 7 \cos(2 \pi y)}{42} e^{-2 \pi x } + {\mathcal O} (e^{-11 \pi x /3})  \right)
\label{openasympangle}
\end{equation}
The first correction term implies that the contours of constant $C(x,y)$ are determined by $x = x_0+ 3/(5 \pi)\log[11 + 5 \cos (2 \pi y)] $ where $x_0$ is a constant.  This is roughly sinusoidal and has an amplitude of $3/(5 \pi)\log(8/3)\approx 0.1873$, which is consistent with the behavior seen in Fig.\ \ref{opencontours}.  Thus, the sinusoidal behavior in the $y$-direction of the correction term for the open b.c.\ persists for all $x$, although its amplitude drops off exponentially in the $x$-direction.

Along the centerline $y = 1/2$, Eq.\ (\ref{openasympangle}) yields
\begin{eqnarray}
C_{\mathrm{open}}(x,1/2) &=& C_0 \left( 1 - \frac{\sqrt{3}\Gamma\left(\frac{1}{3}\right)^6}{40\cdot2^{2/3} \pi^3} e^{-5 \pi x / 3} 
+\frac{1}{7} e^{-2 \pi x } + {\mathcal O} (e^{-11 \pi x /3})  \right) \nonumber \\
&=& C_0 -0.33492472\, e^{-5 \pi x / 3} + 0.14713240 \, e^{-2 \pi x } + {\mathcal O} (e^{-11 \pi x /3})
\label{openasymp}
\end{eqnarray}
In Fig.\ \ref{bobsopen}, we plot $\ln (C_0 - C_\mathrm{open})$ vs.\ $x$, which should approach a straight line with slope $-5 \pi/3$ for 
$0 \ll x \ll w/2$, where $w = 4$ is the width of this system.
The agreement between the prediction and theory is seen to be excellent for $x < 1$.

The first two exponents in (\ref{openasymp}) are close to each other, but far from the next exponent ($-11 \pi x /3$).
In Fig.\ \ref{bobsopen} we also show how the numerical data compares with using those 
first two corrections terms in (\ref{openasymp}).  They give an excellent fit for $1/2 < x < 1$.

\subsection{Rectangle}\label{FinRect}

We now consider the finite rectangle. Here, we determine the five point function in (\ref{PLRz2}), and extend our results to $y > 0$ by assuming the $y$-independence observed numerically.   

The Kac null states and associated differential equations are less helpful in the case of correlation functions with more than four operators.  Instead we find the conformal blocks using vertex operators in a Coulomb gas formalism using the methods described in \cite{DotsenkoFateev84}.  This method, as it relates to four-point function, is fully described in \cite{DotsenkoFateev84}. The generalization to five-point functions is straightforward. 

The central charge of the conformal field theory is $c=0$ so that the Coulomb-gas background charge is $-2 \alpha_0=-2/\sqrt{24}$.  The vertex operators with charge $\alpha$, $V_{\alpha}(z)$, correspond to conformal operators with weight $h=\alpha(\alpha-2\alpha_0)$.  Thus for $\phi_{1,2}$ with $h_{1,2}=0$ we require $\alpha=0$ or $2 \alpha_0$; for $\phi_{1,3}$ with $h_{1,3}=1/3$ we have $\alpha=-2 \alpha_0$ or $4 \alpha_0$.  By the charge neutrality condition only correlation functions with net charge $2 \alpha_0$ are non-zero. In the plane, these take on the form
\be
\langle \prod_i V_{\alpha_i}(z_i) \rangle = \prod_{i<j}(z_j-z_i)^{2 \alpha_i \alpha_j}\; .
\ee

In order to calculate correlation functions consistently we also need to include zero weight screening operators.  These non-local operators are formed by taking the integral of weight $1$ operators around a closed loop; they compensate for any excess charge in the original set of operators.  Since $h=1$ allows either $\alpha=-4 \alpha_0$ or $6 \alpha_0$ we have two screening operators at our disposal.
\be
\oint_{\Gamma}\mathrm{d}u V_{(1\pm5)\alpha_0}(u)\; ,
\ee
where $\Gamma$ is the closed contour of integration, which must non-trivially wrap around the singular points of the correlator.  

For our purposes, in the half-plane coordinates, the simplest expression for (\ref{PLRz2}) with the correct scaling weights and charge neutrality is
\be
\langle \phi_{1,2}( -\infty) \phi_{1,2}(1-1/\mu)\phi_{1,2}(0)\phi_{1,2}(\nu) \phi_{1,3}(1)\rangle=\oint_\Gamma \langle  V_{2 \alpha_0}(-\infty) V_{2 \alpha_0}(1-1/\mu) V_{2 \alpha_0}(0) V_{2 \alpha_0}(\nu) V_{-2 \alpha_0}(1) V_{-4 \alpha_0}(u) \rangle \mathrm{d}u\; ,
\ee
with  $\nu, \mu \in (0,1)$.  The intervals $(-\infty, 1-1/\mu)$, and $(0, \nu)$ are wired, while the point at $1$ is on the free boundary outside of these intervals.  

In most cases the contour $\Gamma$ can be chosen so that it can be replaced with an open contour between singular points of the correlation function \cite{DotsenkoFateev84}.  This means we can rewrite the correlator as an integral 
\be
\left(\frac{\nu (1-\mu)(1-\mu +\nu \mu)}{\mu(1-\nu)}\right)^{1/3} \int_\gamma \left( \frac{u-1}{u(u-\nu)(u-1+1/\mu)}\right)^{2/3}\;\mathrm{d}u\; ,
\ee
with $\gamma$ a segment of the real line between two of the five points.

The various choices for $\gamma$ lead to the following formulae, where in each case we pick the branches of the integrand in order to obtain a real solution:
\begin{eqnarray} \label{WAB2}
( -\infty, 1-1/\mu )=\gamma &\Rightarrow&W_{AB_2\phantom{B_1}}=\frac{\Gamma(1/3)^2}{\Gamma(2/3)}\frac{ \nu^{1/3}}{(1-\nu)^{1/3}(1-\mu +\mu \nu)^{1/3}} \mathrm{F}_1\left( \textstyle \frac{1}{3}; \frac{2}{3},-\frac{2}{3}; \frac{2}{3} \big| \frac{\nu \mu}{1-\mu+\mu \nu},\mu\right)\\ \label{WAB1B2}
(1-1/\mu,0)=\gamma &\Rightarrow&W_{AB_1B_2}=\frac{\Gamma(1/3)^2}{\Gamma(2/3)}\frac{ \nu^{1/3}}{(1-\nu)^{1/3}(1-\mu +\mu \nu)^{1/3}} \mathrm{F}_1\left( \textstyle \frac{1}{3}; \frac{2}{3},-\frac{2}{3}; \frac{2}{3} \big| \frac{1- \mu}{1-\mu+\mu \nu},1-\mu\right)\\ \label{WAB1}
(0,\nu)=\gamma&\Rightarrow&W_{AB_1\phantom{B_2}}=\frac{\Gamma(1/3)^2}{\Gamma(2/3)}\frac{ \mu^{1/3}}{(1-\nu)^{1/3}} \mathrm{F}_1\left( \textstyle \frac{1}{3}; \frac{2}{3},-\frac{2}{3}; \frac{2}{3} \big| \frac{ \mu \nu}{1-\mu+\mu \nu},\frac{\nu}{1-\mu+\mu \nu}\right)\\ \label{WB1}
(\nu, 1 )=\gamma &\Rightarrow&W_{B_1\phantom{AB_2}}=\frac{4 \pi}{\sqrt{27} }\left( \mu \nu (1-\mu)(1-\nu)^2(1-\mu +\mu \nu)\right)^{1/3} \mathrm{F}_1\left( \textstyle \frac{5}{3}; \frac{2}{3},\frac{2}{3}; 2 \big| \mu(1-\nu),1-\nu \right)\\
(1, \infty)=\gamma &\Rightarrow&W_{B_2\phantom{AB_1}}=\frac{4 \pi}{\sqrt{27} }\left( \mu \nu (1-\mu)(1-\nu)^2(1-\mu +\mu \nu)\right)^{1/3} \mathrm{F}_1\left( \textstyle \frac{5}{3}; \frac{2}{3},\frac{2}{3}; 2 \big| \nu,1-\mu +\mu \nu \right) \label{WB2}
\end{eqnarray}

\begin{figure}[htbp]
\begin{center}
 \includegraphics[scale=0.6]{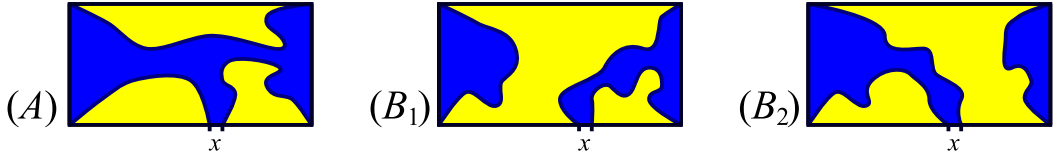}
\caption{The three types of configurations resulting from wired boundary conditions in the rectangle.}
\label{Rect}
\end{center}
\end{figure}

\begin{figure}[htbp]
\begin{center}
\includegraphics[scale=0.6]{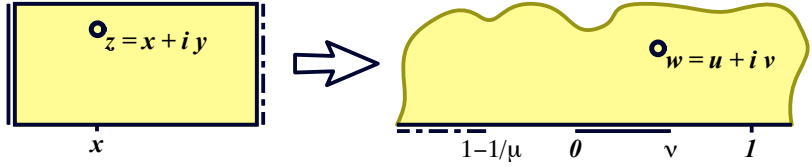}
\caption{Mapping the rectangle onto the upper-half plane: the left side maps onto $(0,\nu)$, the right side maps onto $(-\infty, 1-1/\mu)$ and the point on the bottom side maps to $1$.}
\label{KeyRect}
\end{center}
\end{figure}

These expressions, which involve the Appell hypergeometric function $\mathrm{F}_1(a;b_1,b_2;c|z_1,z_2)$, are conformal blocks of (\ref{PLRz2}).  The blocks represent the configurations in Fig. \ref{Rect} as indicated by their subscripts, i.e., $W_{AB_1B_2}$ represents $A \cup B_1 \cup B_2$, while $W_{B_1}$ represents the $B_1$ configurations only.  Thus, for example, $W_A = W_{AB_1} - W_{B_1} = W_{AB_2} - W_{B_2}$. 

In order to justify our assignments for these conformal blocks we take the limits $\nu$ or $\mu \to 0$ for comparison with those found directly from the five-point function. Taking the leading term of the five-point function as either $\nu$ or $\mu \to 0$ gives the three or four-point functions of section \ref{SIStrip}, which allows straightforward identification with the physical configurations.  

As $\nu \to 0$ the weights of the configurations in Fig. \ref{Rect} are
\bea \label{WAf}
\Omega_A & \to&C_{1 1 2} \nu^{1/3} \langle \phi_{1,2}( -\infty)  \phi_{1,2}(1-1/\mu)_{[1,3]} \phi_{1,3}(0) \phi_{1,3}(1)\rangle = \nu^{1/3} C_{1 1 2}{}^2C_{2 2 2}F_{\phi_{1,3}}(\mu)\;,\\\label{WB1f}
\Omega_{B_1} & \to & \langle \phi_{1,2}( -\infty )  \phi_{1,2}(1-1/\mu)\phi_{1,3}(1)\rangle =  C_{1 1 2}\mu^{1/3}\;,\\\label{WB2f}
\Omega_{B_2} & \to & C_{1 1 2} \nu^{1/3} \langle \phi_{1,3}(1) \phi_{1,2}(1-1/\mu)_{\partial[1,2]} \phi_{1,3}(0) \phi_{1,2}(-\infty)\rangle= \nu^{1/3} C_{1 1 2}K_{\partial \phi_{1,2}}\;G_{\partial\phi_{1,2}}(\mu)\;,
\eea
where the notation $\Omega$ is used to emphasize that these expressions come from the original correlation function and therefore may have a different overall normalization than our conformal blocks.
These limits are easily deduced by inserting the leading order term in the boundary operator product expansion for $\phi_{1,2}( 0 )\phi_{1,2}(\nu)$ into (\ref{PLRz2}) as discussed in section \ref{SIStrip}:  if the interval $(0,\nu)$ (the left side in Fig. \ref{Rect}) is isolated from the other structures the leading contribution will be $\mathbf{1}(0)$, and if it is attached to the other structures the leading fusion term will be $C_{1 1 2} \nu^{1/3} \phi_{1,3}(0)$.

For comparison  we take the limit $\nu \to 0$ in the expressions (\ref{WAB2})-(\ref{WB2}) for our conformal blocks.  We find
\bea \label{WAB2alt}
W_{AB_2}&\to&\frac{\Gamma(1/3)^2}{\Gamma(2/3)}\frac{ \nu^{1/3}}{(1-\mu)^{1/3}} \mathrm{F}_1\left( \textstyle \frac{1}{3}; \frac{2}{3},-\frac{2}{3}; \frac{2}{3} \big| 0 ,\mu\right) =\frac{\Gamma(1/3)^2}{\Gamma(2/3)} \nu^{1/3}F_{\mathbf{1}}(\mu)\;,\quad \mathrm{and}\\ \label{WB2alt}
W_{B_2}&\to&\frac{4 \pi}{\sqrt{27} }\left(\nu \mu (1-\mu)^2 \right)^{1/3} \mathrm{F}_1\left( \textstyle \frac{5}{3}; \frac{2}{3},\frac{2}{3}; 2 \big| 0,1-\mu \right)=\frac{4 \pi}{\sqrt{27} }\nu^{1/3} G_{\partial\phi_{1,2}}(\mu)\;.
\eea
Taking into account the crossing symmetry (\ref{Xsym}) we see that $W_{AB_2} \sim \Omega_A + \Omega_{B_2}$ and $W_{B_2} \sim \Omega_{B_2}$, validating our labels for these two blocks.  We find that  (\ref{WAB1B2}), (\ref{WAB1}), and (\ref{WB1}) are ${\cal O}(\nu ^0)$.  Thus these three blocks must represent the  $B_1$ configurations.  In order to specify the inclusion/exclusion of $A$ or $B_2$ in these blocks we next perform a similar analysis for $\mu \to 0$.

In order to take the limit $\mu \to 0$ we must use the expansion around negative infinity; $\phi_{1,2}(-\xi)\phi_{1,2}(1-1/\mu) \to \mathbf{1}(-\xi)$ or $C_{1 1 2} \mu^{1/3} \xi^{2/3} \phi_{1,3}(-\xi)$ with $\xi \gg 1/\mu$ and eventually $\xi \to \infty$. Here
\bea \label{WAg}
\Omega_A & \to&C_{1 1 2} \mu^{1/3} \xi^{2/3} \langle \phi_{1,3}( -\xi)\phi_{1,2}(0)_{[1,3]}\phi_{1,2}(\nu) \phi_{1,3}(1)\rangle = \mu^{1/3} C_{1 1 2}{}^2 C_{2 2 2}F_{\phi_{1,3}}(\nu)\;,\\\label{WB1g}
\Omega_{B_1} & \to &C_{1 1 2} \mu^{1/3} \xi^{2/3}  \langle \phi_{1,3}( -\xi ) \phi_{1,2}(0)\phi_{1,2}(\nu)_{\partial [1,2]}\phi_{1,3}(1)\rangle =   \mu^{1/3}C_{1 1 2} K_{\partial \phi_{1,2}}\; G_{\partial\phi_{1,2}}(\nu)\;,\\\label{WB2g}
\Omega_{B_2} & \to & \langle\phi_{1,2}(0)\phi_{1,2}(\nu) \phi_{1,3}(1)\rangle= C_{1 1 2}\nu^{1/3} (1-\nu)^{-1/3}\;.
\eea
We then take the limit $\mu \to 0$ in the expressions (\ref{WAB2})-(\ref{WB2}), which yields
\bea \label{WAB1alt}
W_{AB_1}&\to&\frac{\Gamma(1/3)^2}{\Gamma(2/3)}\frac{ \mu^{1/3}}{(1-\nu)^{1/3}} \mathrm{F}_1\left( \textstyle \frac{1}{3}; \frac{2}{3},-\frac{2}{3}; \frac{2}{3} \big| 0 ,\nu\right) =\frac{\Gamma(1/3)^2}{\Gamma(2/3)} \mu^{1/3}F_{\mathbf{1}}(\nu)\;,\quad \mathrm{and}\\ \label{WB1alt}
W_{B_1}&\to&\frac{4 \pi}{\sqrt{27} }\left(\nu \mu (1-\nu)^2 \right)^{1/3} \mathrm{F}_1\left( \textstyle \frac{5}{3}; \frac{2}{3},\frac{2}{3}; 2 \big| 0,1-\nu \right)=\frac{4 \pi}{\sqrt{27} }\mu^{1/3} G_{\partial\phi_{1,2}}(\nu)\;.
\eea
The comparison validates our assignment of $W_{AB_1}$ and $W_{B_1}$.  

Having identified these four blocks is sufficient for our purposes but for completeness we can also justify our label on  the fifth block,  $W_{AB_1B_2}$.  If we deform the contour of integration from $1-1/\mu$ to $0$ into the upper half plane so that it runs just above the real line from $1-1/\mu$ to $-\infty$ and from $\infty$ to $0$ then we can express $W_{AB_1B_2}$ in terms of the other four conformal blocks,
\be
W_{AB_1B_2}=-e^{4 \pi i/3}W_{AB_2}-e^{2 \pi i/3}W_{AB_1}-e^{4 \pi i/3}W_{B_1}-e^{2 \pi i/3}W_{B_2}\; .
\ee 
Taking the real and imaginary parts of this relation gives: $W_{AB_1}+W_{B_2}=W_{AB_2}+W_{B_1}$, a non-trivial check of our established labels, and $W_{AB_1B_2} =\frac{1}{2}\left(W_{AB_1}+W_{B_2}+W_{AB_2}+W_{B_1}\right) $, which validates our final label. 

Now that we have the necessary results in the upper half-plane, we transform them into the rectangle.  The analytic function that takes the rectangle $z=\{x+i y | x\in(0,R),y\in(0,1)\}$ to the upper half-plane mapping $(i,0,x,R,R+i)$ (with $x \in(0,R)$) onto $(0,\nu,1,-\infty,1-1/\mu)$ can be written 
\be
f(z)=\frac{\mathrm{nc}^2(z K(1-a)|a)}{\mathrm{nc}^2(x K(1-a)|a)}\;,
\ee
where $\mathrm{nc}(z|m)$ is a Jacobi elliptic function and $K(m)$ the complete elliptic integral of the first kind.  
We determine the elliptic parameter $a$ from $R$ in a slightly non-standard way.  The usual elliptic nome in this geometry would be $q=e^{-\pi/R}$, but we prefer to expand in terms of $q'=e^{-\pi R}$. Since $R \to 1/R$ is equivalent to $a \to 1-a$ we find that
\be
a=1-\frac{\vartheta_2(0,q')^4}{\vartheta_3(0,q')^4}=\frac{\vartheta_4(0,q')^4}{\vartheta_3(0,q')^4} \; ,
\ee
using the standard expression for the inverse nome and elliptic theta function identities \cite{AbSt}.

Using elliptic theta functions, the undetermined parameters in the mapping are
\bea
\nu &=& f(0)=\mathrm{cn}^{2}(x K(1-a)|a)=\frac{\vartheta_2(0,q')^2\vartheta_4(i \pi x/2,q')^2}{\vartheta_4(0,q')^2\vartheta_2(i \pi x/2,q')^2}\; ,\quad \mathrm{and}\\
\mu&=&\frac{1}{1-f(R+i)} = \frac{1}{1+\frac{a}{1-a}\,\mathrm{cn}^2(x K(1-a)|a)}=\frac{\vartheta_2(0,q')^2\vartheta_2(i \pi x/2,q')^2}{\vartheta_3(0,q')^2\vartheta_3(i \pi x/2,q')^2}\; .
\eea
Thus the probabilities on the rectangle are
\bea
P_{L R}(x) &=&f'(x)^{1/3} W_A \; ,\\
P_L(x)&=&f'(x)^{1/3} W_{AB_2} \; ,\\
P_R(x)&=&f'(x)^{1/3} W_{AB_1} \; ,
\eea
where the conformal transformation introduces the factor
\be 
f'(x)^{1/3}=\left(2K(1-a)\frac{\mathrm{sc}(x K(1-a)|a)}{\mathrm{nd}(x K(1-a)|a)}\right)^{1/3}=\left(-2i K\left(\frac{\vartheta_2(0,q')^4}{\vartheta_3(0,q')^4}\right)\frac{\vartheta_1(i \pi x/2,q')\vartheta_3(i \pi x/2,q')}{\vartheta_2(i \pi x/2,q')\vartheta_4(i \pi x/2,q')}\right)^{1/3}\; .
\ee
We include this for completeness even though it cancels in the ratios that follow.  

We emphasize that these probabilities are not physically normalized, since  certain non-universal constant factors related to the point operator in the correlation function are ignored.  Note also that in arriving at the above expressions we have used the Jacobi elliptic functions cn$(z|m)$, sc$(z|m)$ and nd$(z|m)$, and subsequently replaced them with their elliptic theta function equivalents.

The final expressions for the probabilities of the various configurations are 
\begin{eqnarray} \label{eqone}
P_{A \cup B_2}&=&\frac{\Gamma\left(\frac{1}{3}\right)^2}{\Gamma\left(\frac{2}{3}\right)}\left(\frac{i\vartheta_2(0)^2\vartheta_4(0)^2\vartheta_3\left(\frac{i \pi x}{2}\right)^3}{\vartheta_3(0)^4\vartheta_1\left(\frac{i \pi x}{2}\right)\vartheta_2\left(\frac{i \pi x}{2}\right)\vartheta_4\left(\frac{i \pi x}{2}\right)} \right)^{1/3} \mathrm{F}_1\left( \frac{1}{3}; \frac{2}{3},-\frac{2}{3}; \frac{2}{3} \bigg| \frac{\vartheta_2(0)^4}{\vartheta_3(0)^4},\frac{\vartheta_2(0)^2\vartheta_2\left(\frac{i \pi x}{2}\right)^2}{\vartheta_3(0)^2\vartheta_3\left(\frac{i \pi x}{2}\right)^2}\right) \; , \\
P_{A \cup B_1 \cup B_2}&=&\frac{\Gamma\left(\frac{1}{3}\right)^2}{\Gamma\left(\frac{2}{3}\right)}\left(\frac{i\vartheta_2(0)^2\vartheta_4(0)^2\vartheta_3\left(\frac{i \pi x}{2}\right)^3}{\vartheta_3(0)^4\vartheta_1\left(\frac{i \pi x}{2}\right)\vartheta_2\left(\frac{i \pi x}{2}\right)\vartheta_4\left(\frac{i \pi x}{2}\right)} \right)^{1/3} \mathrm{F}_1\left( \frac{1}{3}; \frac{2}{3},-\frac{2}{3}; \frac{2}{3} \bigg| \frac{\vartheta_4(0)^4}{\vartheta_3(0)^4},\frac{\vartheta_4(0)^2\vartheta_4\left(\frac{i \pi x}{2}\right)^2}{\vartheta_3(0)^2\vartheta_3\left(\frac{i \pi x}{2}\right)^2}\right) \; , \\
&=&\frac{\Gamma\left(\frac{1}{3}\right)^2}{\Gamma\left(\frac{2}{3}\right)}\left(\frac{i\vartheta_2(0)^2\vartheta_4(0)^2\vartheta_2\left(\frac{i \pi x}{2}\right)^3}{\vartheta_3(0)^4\vartheta_1\left(\frac{i \pi x}{2}\right)\vartheta_3\left(\frac{i \pi x}{2}\right)\vartheta_4\left(\frac{i \pi x}{2}\right)} \right)^{1/3} \mathrm{F}_1\left( \frac{1}{3}; \frac{2}{3},-\frac{2}{3}; \frac{2}{3} \bigg| \frac{\vartheta_4(0)^4}{\vartheta_3(0)^4},-\frac{\vartheta_4(0)^2\vartheta_1\left(\frac{i \pi x}{2}\right)^2}{\vartheta_3(0)^2\vartheta_2\left(\frac{i \pi x}{2}\right)^2}\right) \; , \\ \label{eqtwo}
P_{A \cup B_1}&=&\frac{\Gamma\left(\frac{1}{3}\right)^2}{\Gamma\left(\frac{2}{3}\right)}\left(\frac{i\vartheta_2(0)^2\vartheta_4(0)^2\vartheta_2\left(\frac{i \pi x}{2}\right)^3}{\vartheta_3(0)^4\vartheta_1\left(\frac{i \pi x}{2}\right)\vartheta_3\left(\frac{i \pi x}{2}\right)\vartheta_4\left(\frac{i \pi x}{2}\right)} \right)^{1/3} \mathrm{F}_1\left( \frac{1}{3}; \frac{2}{3},-\frac{2}{3}; \frac{2}{3} \bigg| \frac{\vartheta_2(0)^4}{\vartheta_3(0)^4},\frac{\vartheta_2(0)^2\vartheta_3\left(\frac{i \pi x}{2}\right)^2}{\vartheta_3(0)^2\vartheta_2\left(\frac{i \pi x}{2}\right)^2}\right) \; , \\ \label{eqthr}
P_{B_1}&=&\frac{4 \pi}{\sqrt{27} }\left(\frac{\vartheta_2(0)^4\vartheta_3(0)^2\vartheta_1\left(\frac{i \pi x}{2}\right)^5\vartheta_4\left(\frac{i \pi x}{2}\right)^5}{i\vartheta_4(0)^6\vartheta_2\left(\frac{i \pi x}{2}\right)^5\vartheta_3\left(\frac{i \pi x}{2}\right)^5} \right)^{1/3} \mathrm{F}_1\left( \frac{5}{3}; \frac{2}{3},\frac{2}{3};2 \bigg| -\frac{\vartheta_2(0)^2\vartheta_1\left(\frac{i \pi x}{2}\right)^2}{\vartheta_4(0)^2\vartheta_3\left(\frac{i \pi x}{2}\right)^2},-\frac{\vartheta_3(0)^2\vartheta_1\left(\frac{i \pi x}{2}\right)^2}{\vartheta_4(0)^2\vartheta_2\left(\frac{i \pi x}{2}\right)^2}\right) \; ,  \nonumber \\
\\
P_{B_2}&=&\frac{4 \pi}{\sqrt{27} }\left(\frac{\vartheta_2(0)^4\vartheta_3(0)^2\vartheta_1\left(\frac{i \pi x}{2}\right)^5\vartheta_4\left(\frac{i \pi x}{2}\right)^5}{i\vartheta_4(0)^6\vartheta_2\left(\frac{i \pi x}{2}\right)^5\vartheta_3\left(\frac{i \pi x}{2}\right)^5} \right)^{1/3} \mathrm{F}_1\left( \frac{5}{3}; \frac{2}{3},\frac{2}{3};2 \bigg| \frac{\vartheta_2(0)^2\vartheta_4\left(\frac{i \pi x}{2}\right)^2}{\vartheta_4(0)^2\vartheta_2\left(\frac{i \pi x}{2}\right)^2},\frac{\vartheta_3(0)^2\vartheta_4\left(\frac{i \pi x}{2}\right)^2}{\vartheta_4(0)^2\vartheta_3\left(\frac{i \pi x}{2}\right)^2}\right) \; .
\end{eqnarray}
Combining (\ref{eqone}), (\ref{eqtwo}), (\ref{eqthr}) and including the horizontal crossing probability
\be
\Pi_h = \frac{2 \pi 3\sqrt{3}}{\Gamma\left( \frac{1}{3}\right)^3}\left(\frac{\vartheta_2(0)^4}{\vartheta_3(0)^4}\right)^{1/3}{}_2F_1\left(\frac{1}{3},\frac{2}{3};\frac{4}{3};\frac{\vartheta_2(0)^4}{\vartheta_3(0)^4}\right)
\ee
given by Cardy's formula with cross-ratio $\mu \nu/(1-\mu +\mu \nu)$, we find the main result of this section
\be \label{Crect}
C(z)=C(x)=\frac{P_{L R}(x)}{\sqrt{P_L(x)P_R(x)\Pi_h}}=\frac{P_{A \cup B_1}-P_{B_1}}{\sqrt{P_{A \cup B_1}P_{A \cup B_2}\Pi_h}} \; .
\ee
This result is compared with simulations in the next section.  

Note that our results give exact expressions for $P_{L }(x)$, $P_{ R}(x)$, and $P_{L R}(x)$ in an arbitrary rectangle.

\subsection{Comparison with numerical results for a rectangle} \label{comps}

\begin{figure}
\begin{center}
\includegraphics[width=3in]{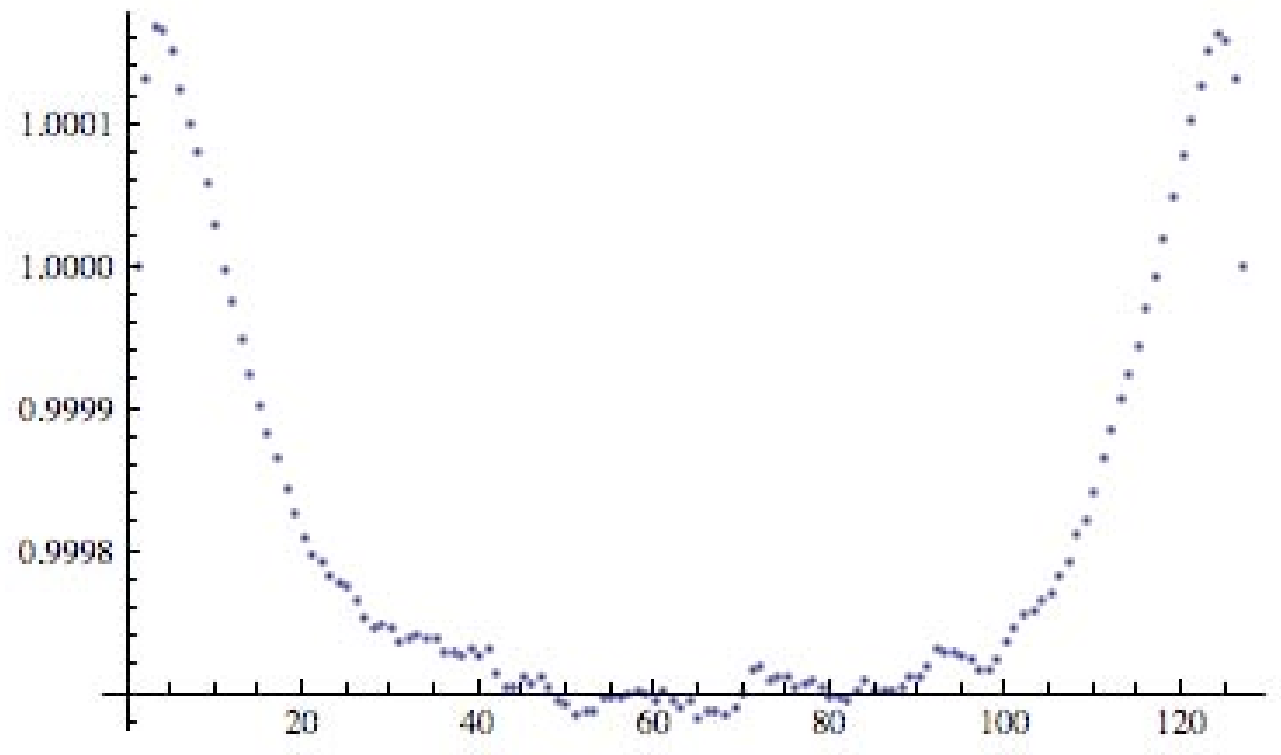}
\includegraphics[width=3in]{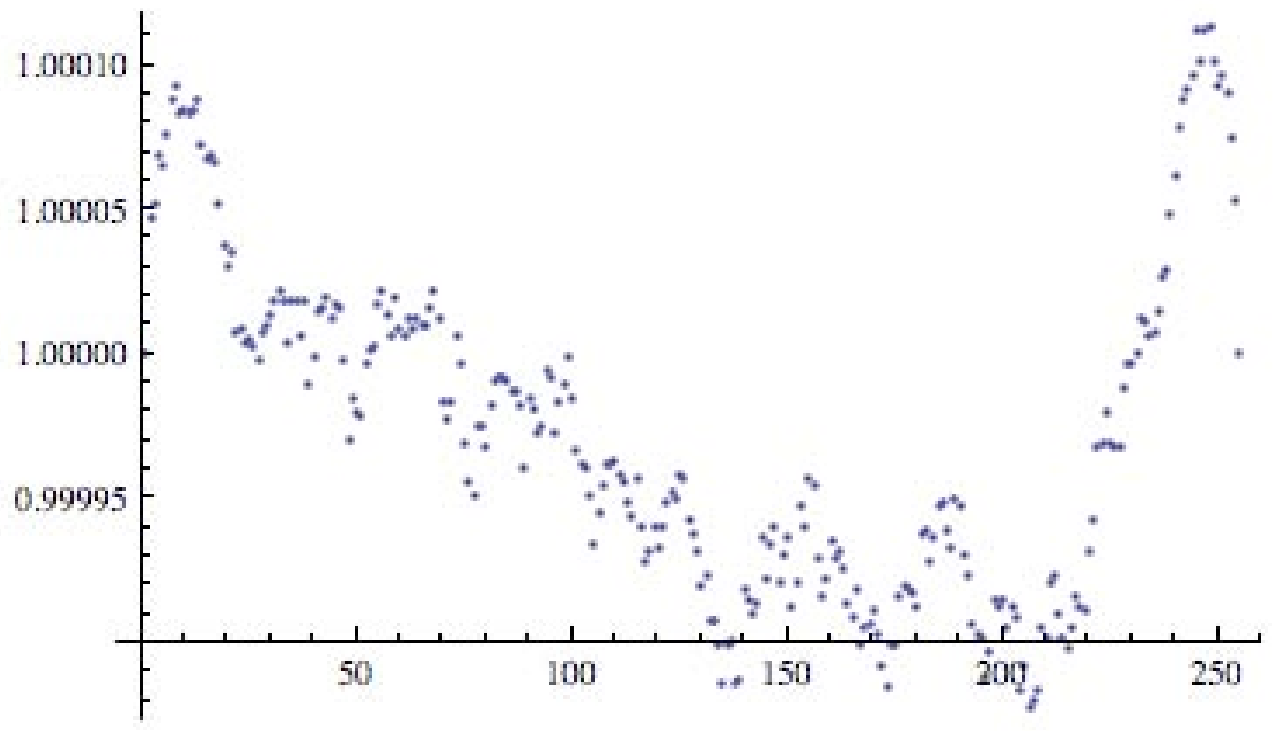}
\caption{Ratio of measurements of $C(z)$ (averaged over all $y$) to theory (\ref{Crect}) for bond percolation systems of size $63 \times 127$ (left) and $127 \times 255$ (right) with wired b.c., plotted as a function of the lattice coordinate $X$.}
\label{ratiofig}
\end{center}
\end{figure}

In Fig.\ \ref{ratiofig} we show the ratio of the measured values of $C(x)$ to the theory, Eq.\ (\ref{Crect}), for systems of size $63 \times 127$ and $127 \times 255$.  There is a slight overshoot near the boundaries, which was lowered to some extent by assuming that the effective boundaries of the system were 0.2 lattice spacings inside the system.  There is a  small undervalue in the center which decreases as the system size increases.

\begin{figure}
\begin{center}
\includegraphics[width=4in]{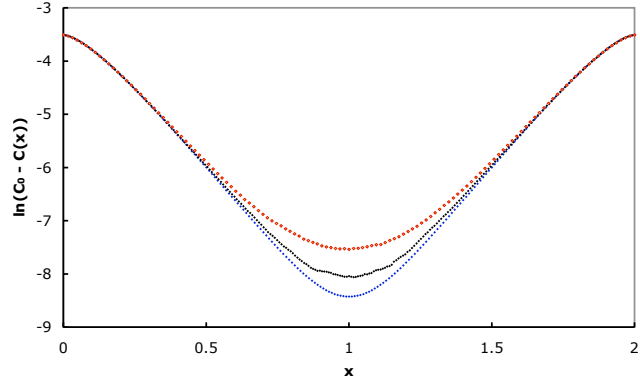}
\caption{$\ln[C_0 - C(x)]$ vs.\ $x$, wired b.c., for systems of size $63 \times 127$ (top) and $127 \times 255$ (center), and theory for a system of aspect ratio 2 (bottom).}
\label{rectfig}
\end{center}
\end{figure}

Another way to illustrate the difference is shown in Fig.\ \ref{rectfig}, in which we plot $\ln (C_0 - C(x))$ vs.\ $x$ for a systems of aspect ratio 2 for two different sizes.  It can be seen that as the size increases, the deviations from theory around $x = 1$ decrease.  Thus, overall, we find very good agreement between Eq.\ (\ref{Crect}) and simulations.

\begin{figure}
\begin{center}
\includegraphics[width=5in]{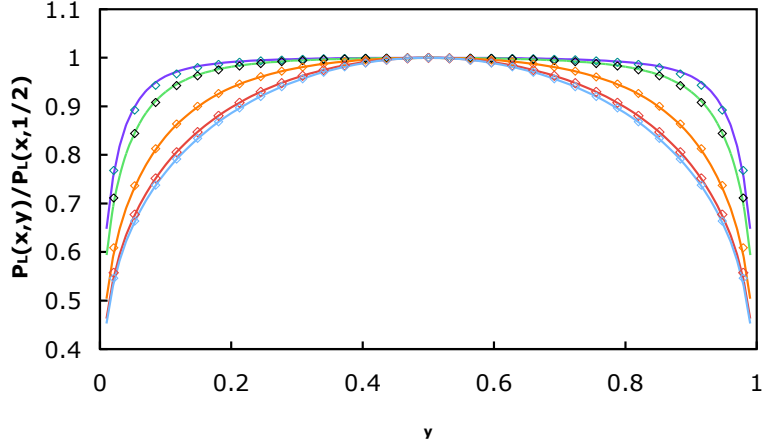}
\caption{Numerical values of  $P_L(x,y)/P_L(x,1/2)$ as a function of vertical coordinate $y$ for $X = 3, 4, 8, 16, 32$ (top to bottom) on a rectangular lattice of size $31 \times 63$, with wired vertical boundaries (data points).  We set $x = (X - 1)/62$ and $y = (Y - 1)/30$.  The corresponding results for $P_R$ and $P_{LR}$ are identical, as are the results for $x$ replaced by $1 - x$.  The curves are drawn assuming the strip results (\ref{PLw}),  which are are seen to describe the rectangular results quite closely.}
\label{ydependence}
\end{center}
\end{figure}

In Fig.\ \ref{ydependence} we explore the question of the vertical dependence of the correlation functions.   In section \ref{nos} we observed  that the various functions $P_L(z)$, $P_R(z)$, $P_{L R}(z)$ all have the same $y$-dependence, given by Eq.\ (\ref{PLw}) for the case of a strip but unknown for a rectangle.  In Fig.\ \ref{ydependence}  we plot the measured values of $P_L(x,y)/P_L(x,1/2)$ as a function of $y$ for fixed values of $x$.  Here we are normalizing by the value half-way up, $P(x, 1/2)$; we find that the strip result (\ref{PLw}) is indeed quite accurate here.  The results for the other functions  $P_R(z)$, $P_{L R}(z)$ are identical. 

Note that the system displays an interesting horizontal symmetry.  The $y$ dependence of, for example, $P_L(z)$ shown in the top figure of Fig.\ \ref{wiredfig} at $x$ and $W - x$ is the same. Thus since $P_R(x,y) = P_L(W-x,y)$ this implies that  $P_L(z)/P_R(z)$ is only a function of $x$, as discussed above.

\begin{figure}
\begin{center}
\includegraphics[width=2in]{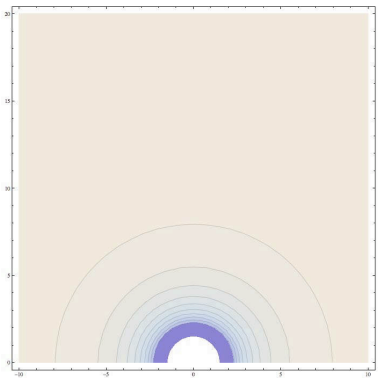}
\caption{Theoretical density plots of $C(z)$ around a semicircular anchor of radius $\e = 1$ in a half-plane from Eq.\  (\ref{FScorr}).  Contours are at $1.029$, $1.028$, $1.027 \ldots$ going inward.}
 \label{theoranchordens}
\end{center}
\end{figure}

\begin{figure}
\begin{center}
\includegraphics[width=2.5in]{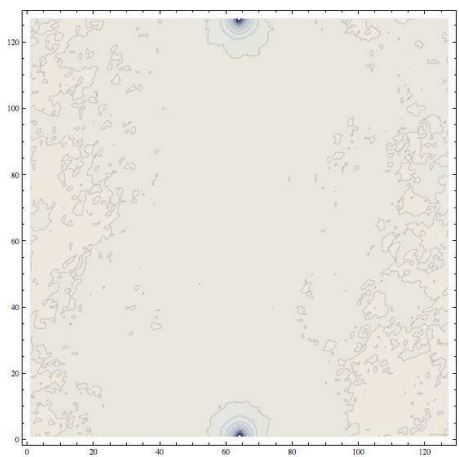}
\caption{Measured density contours of $C(z)$ on a lattice system of size $ 127 \time 127$ with anchors at the top and bottom,
for bond percolation.  Contours around the anchors (going inward) are at $1.029$, $1.028$, $1.027 \ldots$; scattered contours near left and right sides are at $C = 1.03$.}
\label{anchorplot}
\end{center}
\end{figure}

\section{Finite size corrections around an anchor point}\label{FScorrs}

In this section, we use some of the results obtained above to explain  finite-size corrections observed in a different but related problem.  In \cite{KlebanSimmonsZiff06} we examined factorization of correlations in rectangles (or the half-plane) that are  anchored at {\it points} on the boundary. The result is closely analogous to (\ref{Cratio}), except that the clusters are conditioned to touch specified points on the boundary rather than intervals. Here, everywhere except near the anchor points, $C$ takes on, as mentioned, the same constant value $C_0$ found above.  Further, $C=1$ at the anchor points only, and most of the deviation of $C$ from  its asymptotic value  occurs within of the order of $10$ lattice spacings of the anchors.   Since the deviations are only significant for a finite number of lattice spacings, they are a finite-size effect and vanish in the field theory limit.

We can nonetheless use our conformal field theory results to model these corrections to $C$ near the anchor points.  We model the boundary anchor point as a semi-circle of radius $\e$ centered on the origin in the upper-half plane and place the other anchoring point at infinity.  This mimics the case where the anchoring point has an effective finite size that is much smaller than its distance from other significant points and boundary features.  The mapping $w(z) = \e e^{\pi z}$ from the semi-infinite strip to the model geometry implies the relation $\e/r = e^{-\pi x}$ between the horizontal strip coordinate, $x$, and the radial coordinate $r \in \{\e,\infty\}$ of the target space.   Inserting this relation into (\ref{ApproxSIStrip}) we find
\be \label{FScorr}
C(r) = C_0 \; \frac{{}_2F_1(  -\frac12,-\frac13,\frac76, \frac{\e^2}{r^2} )}{\sqrt{{}_2F_1( -\frac12, -\frac23, \frac56  , \frac{\e^2}{r^2})}} \; ,
\ee
with $C$ independent of $\theta$ (the contours are semi-circles).
For large $r$, one has (similar to (\ref{ApproxSIStripAsymp}))
\be
\label{FScorrAsymp}
C(r) = C_0 \left(1 - \frac{2}{35}  \frac{\e^2}{r^2} + \frac{834}{25025} \frac{\e^4}{r^4} -\frac{6406}{734825} \frac{\e^6}{r^6}\ldots \right) \; .
\ee
so that the finite-size corrections drop off to first order as $1/r^2$.  This prediction is illustrated in Fig.\ \ref{theoranchordens}.

\begin{figure}
\begin{center}
\includegraphics[width=6in]{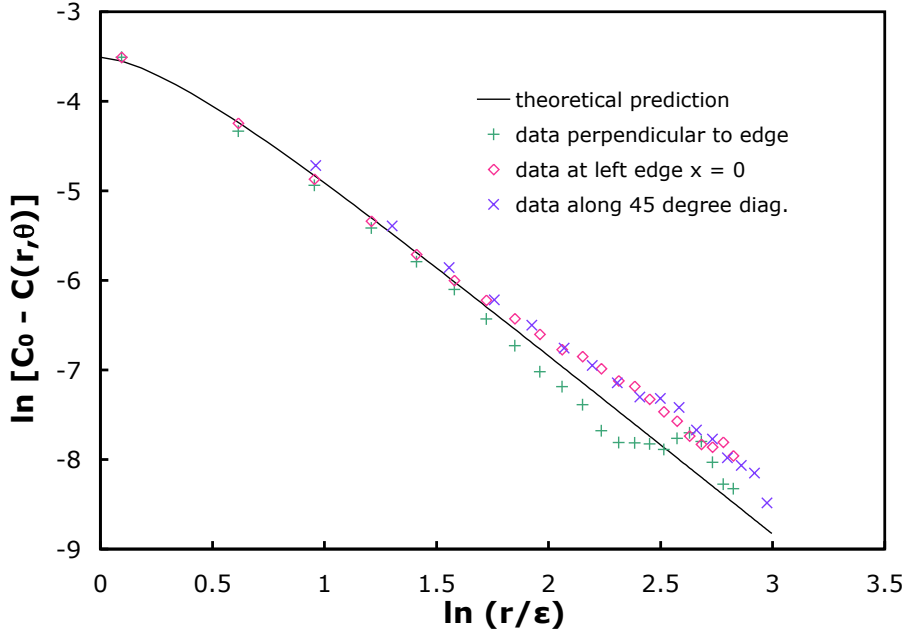}
\caption{ Comparison of  (\ref{FScorr}) with simulations for $C(r,\theta)$ along three radial directions $\theta$ away from an anchor point, where $r$ is the radial distance, and $\varepsilon = 1.34$ lattice spacings.}\label{FSfig}
\end{center}
\end{figure}

To test this prediction, we measured $C(z)$ for a system of $127 \times 127$ lattice spacings with the anchors at the centers of opposite sides, and free boundary conditions elsewhere, for critical bond percolation.  The contours are shown in Fig.\ \ref{anchorplot}.  In Fig.\ \ref{FSfig} we compare $\ln[C_0 - C(z)]$ as a function of distance from the anchor point with  (\ref{FScorr}) in the three directions indicated.  The numerical data is scaled so that $\varepsilon = 1.34$ lattice spacings in order to match the theoretical predication.  There is an additional scale factor related to where to put the first point (the boundary location);  that is chosen at $1.15$ radii.  These two scale factors have considerable leeway -- for example, choosing $1.2$ for $\varepsilon$ gives a very good fit for the perpendicular $C(r)$, but not as good for the parallel (side) $C(r)$.  There is more noise along the side ($x=0$), because the probabilities are lower there.  Also, as $r$ increases, $C_0 - C(r)$ decreases, so the noise grows.  For the  $45\degree$ case the two scale factors are multiplied by $\sqrt2$ to be consistent.

Note that, by the conformal transformation above, $\ln r$ in Fig.\ \ref{FSfig} corresponds to $x$ in Fig.\ \ref{rectfig},  for small $x$,  since for small $x$ the rectangle may be approximated by a strip.

Thus, the overall agreement with theory is good, and the extent of measurable finite-size effects (of the order of 10 lattice spacings) agrees with our previous observations.

\begin{figure}
\begin{center}
\includegraphics[width=6in]{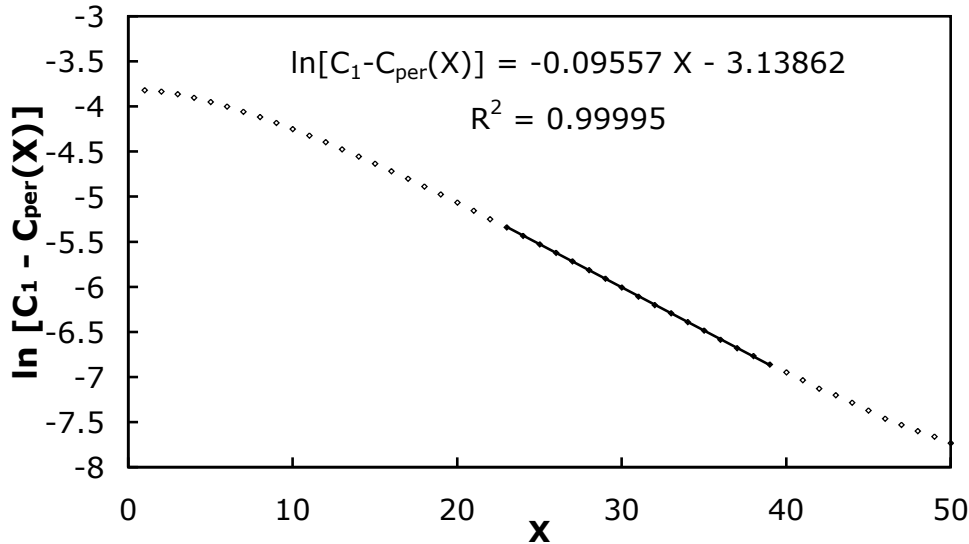}
\caption{Plot of $\ln[C_1- C_\mathrm{per}(x)]$ for a periodic system of size $64 \times 127$, bond percolation, with $C_1 = 1.022$, for $X = 1, \ldots, 50$.  The equation for a linear fit is shown in the figure.} \label{Cperiodic}
\end{center}
\end{figure}

\section{Periodic system} \label{periodicsec}

Next we consider a system with periodic b.c.\ on the horizontal sides 
and wired b.c.\ on the vertical sides.  In this case we have no theoretical predictions for the factorization but just
 present the simulation results, for a system of size $64 \times 127$.   We find that
away from the boundary, $C_\mathrm{per}(x)$ approaches a constant $C_1 \approx 1.022$.  Further, $C_\mathrm{per}(x)$ does not deviate from $C_1$ by more than $2.2 \%$, so factorization is a good approximation everywhere in the system. 

We determined the  value of $C_1$
by plotting $\ln (C_1 - C_\mathrm{per})$ vs.\ $x$ and adjusting $C_1$ to get a good linear fit in the
intermediate regime between the vertical walls and the center at $x = 64$.  This plot is shown in 
Fig.\ \ref{Cperiodic}.   The slope of that fit $\approx -0.09557$ is reasonably close to
the value $-2 \pi / 64 = -0.09817477$ one would get from the expected leading behavior 
$\exp(-2 \pi (X-1) / 64)$.  Assuming higher-order terms are powers of this exponential term, in analogy
with the case of wired b.c., 
we plot $C_\mathrm{per}(x)$ vs.\ $s = \exp(-2 \pi (X-1) / 64)$ in Fig.\ \ref{CvsExpBoth} (lower curve).  
We get an excellent fit to a polynomial with the result
$C_\mathrm{per}(x) = 1.02200(1 - 0.042758 s + 0.014994 s^2 \ldots)$.
Thus, while the behavior of the periodic system is quantitatively different from that of the
open b.c.\ case, qualitatively, the two are quite similar.

\section{Conclusions} \label{cons}

In this paper, we consider two-dimensional systems at the percolation point. We investigate correlation functions between a point in the bulk of a rectangular system and one or both of the two vertical boundaries.  The b.c.\ on  the horizontal sides are assumed to be open (except in the last section where we consider periodic b.c.), and the vertical sides have either wired (fixed) or open b.c.  We discover numerically that for wired b.c.\ on the vertical sides, all the correlation functions have an identical $y$-dependence, which implies that the ratio $C(z)$ (see (\ref{Cratio})) is independent of $y$.  Further, this ratio is close to $1$ for all $z$ in the rectangle, so that factorization of the correlation functions is almost exact everywhere. 

Assuming $y$-independence, we  use boundary operators in conformal field theory to find $C(z)$ and the correlations that comprise it everywhere in the semi-infinite strip (long rectangle)  from the much simpler calculation of the density at the wall.  Explicitly, we solve the differential equations that arise via conformal field theory when the bulk point has been moved to the side of the strip, then use the observed $y$-independence to extend our results away from the boundary.    This is accomplished for both wired (fixed) and open (free) boundary conditions on the vertical ends.

For the rectangle, we solve the density at the wall case with wired boundary conditions by use of vertex operator techniques.  This derivation leads to expressions for the ratios of the various densities, including closed forms for the densities themselves on the boundary.

In the limiting case of the semi-infinite strip we push the conformal calculation a bit further and obtain explicit formulas for all the various probability densities without the assumption of  $y$-independence, verifying the results just mentioned. This is accomplished for both wired (fixed) and open (free) boundary conditions on the vertical ends.   

All of our theoretical results agree very well with the numerical simulations.  Neither finite-size effects nor correction-to-scaling operators make a significant numerical contribution.  This was also observed in related work \cite{SimmonsKlebanZiffPRE07}, which includes comments on this situation.

Taking the limit of a half-infinite strip, and then transforming the left-hand side to a semi-circle, gives a prediction for the behavior around a point as studied in \cite{KlebanSimmonsZiff06}, where the radius $\e$ of the semi-circle is of the order of the lattice spacing.  Thus, using our conformal field theory results, we are able to understand the finite-size effects for that problem, finding for example that the asymptotic behavior $C_0$ is approached as $(\e/r)^2$ where $r$ is the radial distance to the point.

With open boundary conditions on the left- and right-hand sides of the rectangle, the behavior of $C(z)$ is more complicated, and the factorization breaks down near the vertical sides.  However, away from the sides, the function still approaches a constant $C_0$, so there is factorization.  Finally, with periodic boundary conditions on the horizontal sides and wired boundary conditions on the vertical ends, our simulations show approximate factorization everywhere, but  a different asymptotic value for the constant: $C_1 \approx 1.022$.
 
 \section{Acknowledgments}

We dedicate this work to the memory of Oded Schramm, who showed a considerable interest in it.

This work was supported in part by the National Science Foundation Grants Nos.   DMR-0536927 (PK) and  DMS-0553487 (RMZ) and by EPSRC Grant No.\ EP/D070643/1 (JJHS).  PK also acknowledges the hospitality of the Kavli Institute for Theoretical Physics, where part of this work was done, supported by the NSF under Grant No. PHY05-51164.

\appendix*

\section{Full derivation of semi-infinite strip densities.} \label{appendix}

In this appendix, we verify that the solutions for the densities in section \ref{SIStripDens} indeed satisfy the null-state conditions imposed by the $\phi_{1,2}$ bcc operators.  This proves that the $y$-independence is exact.  

Now the correlation functions representing the quantities of interest are of the form
\be
\langle \phi_{1,2}(0) \phi_{1,2}(i) \phi_{1/2,0}(z,\bar z) \phi_{1,3}(\infty) \rangle_{\mathbb{S}}\; ,
\ee evaluated in the semi-infinite strip $\mathbb{S}$.

Consder the analogous quantity in the half-plane $\mathbb{H}$:
\be 
\langle \phi_{1,2}(u_1) \phi_{1,2}(u_2) \phi_{1/2,0}(w,\bar w) \phi_{1,3}(u_3) \rangle_{\mathbb{H}}\; .
\ee
Conformal invariance implies that this correlation function may be written
\be \label{Confform}
(w-\bar w)^{-5/48}\left( \frac{u_2-u_1}{(u_3-u_1)(u_3-u_2)}\right)^{1/3} F\left( \frac{(w-u_1)(u_3-u_2)}{(u_2-u_1)(u_3-w)}, \frac{(\bar w-u_1)(u_3-u_2)}{(u_2-u_1)(u_3-\bar w)} \right)\; ,
\ee
where $F(\eta,\bar \eta)$ is annihilated by the null state differential operators 
\bea 
&&\frac{5/48}{(w-u_1)^2}+\frac{5/48}{(\bar w-u_1)^2}+\frac{2/3}{(u_3-u_1)^2}-\frac{2 \partial_w}{w-u_1}-\frac{2 \partial_{\bar w}}{\bar w-u_1}-\frac{2 \partial_{u_2}}{u_2-u_1}-\frac{2 \partial_{u_3}}{u_3-u_1}-3\partial_{u_1}{}^2\quad \mathrm{and} \\
&&\frac{5/48}{(w-u_2)^2}+\frac{5/48}{(\bar w-u_2)^2}+\frac{2/3}{(u_3-u_2)^2}-\frac{2\partial_w}{w-u_2}-\frac{2\partial_{\bar w}}{\bar w-u_2}+\frac{2\partial_{u_1}}{u_2-u_1}-\frac{2\partial_{u_3}}{u_3-u_2}-3\partial_{u_2}{}^2\; ,
\eea
that arise from the $\phi_{1,2}$ operator at $u_1$ and $u_2$ respectively.

Next apply these two differential operators to (\ref{Confform}) and take the limit $\{ u_1, u_2, u_3 \} \to \{0,1,\infty\}$.   The correlation vanishes as $u_3$ goes to infinity.  We account for this by replacing the second factor in (\ref{Confform}) with the corresponding three-point function.  Thus
\be
\langle \phi_{1,2}(0) \phi_{1,2}(1) \phi_{1/2,0}(w,\bar w) \phi_{1,3}(\infty) \rangle_{\mathcal{H}}=\langle \phi_{1,2}(0) \phi_{1,2}(1) \phi_{1,3}(\infty) \rangle(w-\bar w)^{-5/48} F\left( w, \bar w \right)\; ,
\ee
with $F(w,\bar w)$ satisfying
\bea  \label {DE1}
0 &=& \frac{5(w-\bar w)^2}{48w^2 \bar w^2}F-\frac{2(1-w)^2}{w}\frac{\partial F}{\partial w}-\frac{2(1-\bar w)^2}{\bar w}\frac{\partial F}{\partial \bar w}-3(1-w)^2 \frac{\partial^2 F}{\partial w{}^2}\\ \nonumber
&&\qquad \qquad \qquad \qquad \qquad \qquad -6(1-w)(1-\bar w)\frac{\partial^2 F}{\partial w \partial \bar w}-3(1-\bar w)^2 \frac{\partial^2 F}{\partial \bar w{}^2}\quad \mathrm{and\,,}\\ \label{DE2}
0 &=& \frac{5(w-\bar w)^2}{48(1-w)^2 (1-\bar w)^2}F+\frac{2 w^2}{1-w}\frac{\partial F}{\partial w}+\frac{2 \bar w{}^2}{1-\bar w}\frac{\partial F}{\partial \bar w}-3 w^2 \frac{\partial^2 F}{\partial w{}^2}-6w \bar w\frac{\partial^2 F}{\partial w \partial \bar w}-3 \bar w{}^2 \frac{\partial^2 F}{\partial \bar w{}^2}\; .
\eea
Notice that these equations are interchanged by the transformation $w \to 1-\bar w$, $\bar w \to 1- w$, i.e., there is mirror symmetry about the line $u=1/2$.

We transform these results from complex cooridinates in $\mathbb{H}$ into Cartesian coordinates in $\mathbb{S}$ using the mapping 
\be
w(z) =\frac{\cosh(\pi z)+1}{2}\; ,
\ee
so that
\bea \nonumber
\langle \phi_{1,2}(0) \phi_{1,2}(1) \phi_{1/2,0}(z,\bar z) \phi_{1,3}(\infty) \rangle_{\mathbb{S}}&=&\left(\frac{\partial w}{\partial z}\frac{\partial \bar w}{\partial \bar z}\right)^{5/96} (w(z)-\bar w(\bar z))^{-5/48} F(w(z),\bar w(\bar z))\\ \label{Hfunc}
&=&\left(\frac{\sinh(\pi x)^2+\sin(\pi y)^2}{\sinh(\pi x)^2\sin(\pi y)^2}\right)^{5/96}H(x,y)\; ,
\eea
where $H(x,y) := F\left( w(x+ i y),\bar w(x-iy) \right)$.   Using the relations
\bea
\frac{\partial}{\partial w} &=& \frac{2}{\pi \sinh(\pi z)} \frac{\partial}{\partial z}= \frac{1}{\pi \sinh(\pi (x+iy))} \left( \frac{\partial}{\partial x}-i  \frac{\partial}{\partial y} \right)\quad \mathrm{and}\\
\frac{\partial}{\partial \bar w} &=& \frac{2}{\pi \sinh(\pi \bar z)} \frac{\partial}{\partial \bar z}= \frac{1}{\pi \sinh(\pi (x-iy))} \left( \frac{\partial}{\partial x}+i  \frac{\partial}{\partial y} \right)
\eea
we transform equation (\ref{DE1}) for $F(w,\bar w)$ into one for $H(x,y)$.
Simplifying  yields
\bea \nonumber
0&=& \frac{3 \sinh(\pi x)^2}{\pi^2 (\cos(\pi y)+\cosh(\pi x))^2}\frac{\partial^2H}{\partial x^2}+\frac{6 \sinh(\pi x)\sin(\pi y)}{\pi^2 (\cos(\pi y)+\cosh(\pi x))^2}\frac{\partial^2H}{\partial x \partial y}\\
&&+\frac{3 \sin(\pi y)^2}{\pi^2 (\cos(\pi y)+\cosh(\pi x))^2}\frac{\partial^2H}{\partial y^2}-\frac{(2+\cosh(\pi x)(3\cos(\pi y)+\cosh(\pi x)))\sinh(\pi x)}{\pi (\cos(\pi y)+\cosh(\pi x))^3}\frac{\partial H}{\partial x}\\ \nonumber
&&-\frac{(2+\cos(\pi y)(\cos(\pi y)+3\cosh(\pi x)))\sin(\pi y)}{\pi (\cos(\pi y)+\cosh(\pi x))^3}\frac{\partial H}{\partial y}+\frac{5 \sin(\pi y)^2 \sinh(\pi x)^2}{3(\cos(\pi y)+\cosh(\pi x))^4}H\; .
\eea
The other equation implied by (\ref{DE2}) follows from this one by the mirror symmetry $y \to 1-y$.

\begin{figure}[htb] 
   \centering
   \includegraphics[width=6in]{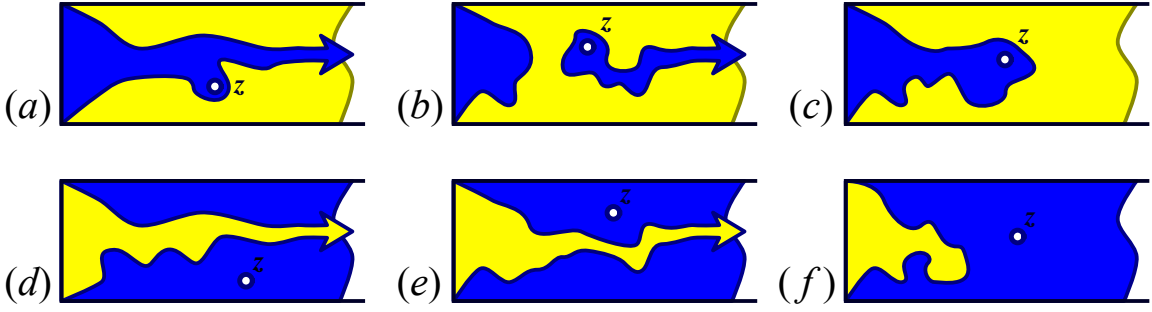} 
   \caption{Configurations in the semi--infinite strip: (\emph{a}) $P_{L R}(z)$, (\emph{b}) $P_{\bar L R}(z)$, (\emph{c}) $P_{L \bar R}(z)$, (\emph{d}) $P_{\bar T B}(z)$, (\emph{e}) $P_{T \bar B}(z)$, (\emph{f}) $P_{T B}(z)$.}
   \label{Conf}
\end{figure}

Now the numerical evidence shows the three densities $P_L(z),P_R(z)$ and $P_{L R}(z)$, that we calculate all take on the form
\be \label{facform}
P_i(z)=Q_i(x)f(x,y),
\ee
with common function $f(x,y)$.  The function $P_L(z)$ (see (\ref{PLw}) and Fig.\ref{Conf}\emph{c}) may be written 

\bea
\langle \phi_{1,2}(0) \phi_{1,2}(i) \phi_{1/2,0}(z,\bar z) \rangle_{\mathcal{S}}&=&\left(\frac{\sinh(\pi x)^2+\sin(\pi y)^2}{\sinh(\pi x)^2\sin(\pi y)^2} \right)^{5/96} \left(\frac{\sin(\pi y)^2}{\sinh(\pi x)^2+\sin(\pi y)^2}\right)^{1/6}  \\
&=&  \frac1{\sinh(\pi x)^{1/3}}  \left(\frac{\sinh(\pi x)^2\sin(\pi y)^2}{\sinh(\pi x)^2+\sin(\pi y)^2}\right)^{11/96} \nonumber  \; .
\eea
This suggests setting
\be \label{fdef}
f(x,y) = \left(\frac{\sinh(\pi x)^2 \sin(\pi y)^2}{\sinh(\pi x)^2+
\sin(\pi y)^2}\right)^{11/96}\; .
\ee
Comparing this to (\ref{Hfunc}) further suggests that we let
\be\label{HtoJ}
H(x,y)=\left(\frac{\sinh(\pi x)^2 \sin(\pi y)^2}{\sinh(\pi x)^2+
\sin(\pi y)^2}\right)^{1/6} J(x,y)\; .
\ee
It follows that if $P_L(z),P_R(z)$ and $P_{LR}(z)$ do have a common $y-$dependence, then there must be solutions for $J(x,y)$ that are independent of $y$ and can be identified with $Q_R(x)$ and $Q_{L R}(x)$. We now demonstrate that this is indeed the case.

Some algebra now shows that     
\bea \label{Eq1}
0&=&-18 \sinh^2(\pi x) \frac{\partial^2 J}{\partial x^2}+3 \pi \sinh(2 \pi x)\frac{\partial J}{\partial x}-18 \sin^2(\pi y) \frac{\partial^2 J}{\partial y^2}+3 \pi \sin(2\pi y)\frac{\partial J}{\partial y} +10 \pi^2 J\; ,\\ \label{Eq2}
0&=&\frac{\partial^2 J}{\partial x \partial y}\; .
\eea
These equations are surprisingly simple, and have definite parity under $y \to 1-y$.

Now (\ref{Eq2}) implies a solution
\be
J(x,y)=g_1(x)+g_2(y)\; ,
\ee
an encouraging result since we expect solutions of precisely this form.  

We insert this  into (\ref{Eq1}) via
\be
J(x,y)=g_3(e^{-2 \pi x})+g_4(\sin(\pi y))
\ee
and express the result in terms of the new variables $X:=e^{-2 \pi x}$ and $Y:= \sin(\pi y)$.  This leads to
\bea \label{diffeq}
0&=&\left(10 g_3(X)-3(1-X)(7-5X)g_3{}'(X)-18X(1-X)^2 g_3{}''(X)\right)\\ \nonumber
&&+\left(10 g_4(Y)+6 Y(1+2Y^2)g_4{}'(Y)-18y^2(1-Y^2)g_4{}''(Y)\right)\; .
\eea

Because $X$ and $Y$ are independent, the two terms in (\ref{diffeq}) must either equal constants that sum to zero (which is the trivial solution since in that case $J(x,y)=0$) or they must be independently equal to zero.  This means that there are four linear solutions for $J(x,y)$ that can be determined from two second order equations.

The two solutions to the $g_3(X)$ equation are
\bea \label{QR}
Q_R(z)&=&\Pi_h \sinh(\pi x)^{-1/3}\, e^{2 \pi x/3}{}_2F_1 \left(-2/3,-1/2,5/6,e^{-2\pi x}\right)\quad \mathrm{and}\\ \label{QLR}
Q_{L R}(z)&=&\Pi_h \sinh(\pi x)^{-1/3}\, e^{\pi x/3}{}_2F_1 \left(-1/2,-1/3,7/6,e^{-2\pi x}\right)\; ,
\eea
which reproduce our expressions for $P_R(z)$ and $P_{L R}(z)$ when combined with the prefactor $f(x,y)$.
For completeness we note that, as mentioned,  $Q_L(z)=\sinh(\pi x)^{-1/3}$ with our particular choice of $f(x,y)$.
This completes the CFT proof of our observation that ratios of $P_L(z)$, $P_R(z)$ and $P_{L R}(z)$ are independent of $y$.

Now this set of conformal operators also yields an additional set of correlation functions.  Logically we expect that these solutions might be some of the corresponding quantities from the analogous limit of vertical crossing in an infinitesimally short rectangle.  The two solutions to the $g_4(Y)$ equation are given by
\bea \label{extr1}
Q_{T \bar B}(z)&=&\Pi_h \sin^{5/3}(\pi y) {}_2F_1\left(4/3,5/3,2,\sin^2\left(\frac{\pi y}{2} \right) \right)\;\quad \mathrm{and}\\  \label{extr2}
Q_{\bar T B}(z)&=&\Pi_h \sin^{5/3}(\pi y) {}_2F_1\left(4/3,5/3,2,\cos^2\left(\frac{\pi y}{2} \right) \right)\;
\eea
which represent configurations with the top and bottom conditioned not to belong to the same cluster, and the bulk point correlated with the top (Fig.\ \ref{Conf}\emph{e})  and bottom (Fig.\ \ref{Conf}\emph{d}) sides respectively as in Fig.\ \ref{Conf}.  

These associations are based on leading term behavior as $y \to 0$ and $1$.  For example, $P_{T \bar B}(z) \sim y^2$ and $(1-y)^0$.  The weight $0$ corresponds to the identity operator, which occurs in the bulk boundary OPE when the bulk operator approaches a fixed interval with which it is correlated.  The weight $2=h_{1,5}$ corresponds to the Fourtuin-Kastelyn four-leg operator, which we expect to appear when the bulk operator approaches a fixed boundary of different spin.  This is best understood as the bulk operator pinching a dual cluster between itself and the boundary, a necessary condition in order that the boundary and bulk point have different spins.  The inner and outer side of the dual cluster emanate from the boundary on both sides of the bulk point leading to a total of four Fourtuin-Kastelyn hulls.  The combination of these two limits uniquely associates (\ref{extr1}) with Fig.\ \ref{Conf}\emph{e} while an identical argument fixes the association of (\ref{extr2}) with Fig.\ \ref{Conf}\emph{d}.

For completeness we include
\be 
Q_{T B}(z) = \sin(\pi y)^{-1/3}
\ee
which follows from the correlation function $\langle \phi_{1,2}(i)\phi_{1,2}(0) \phi_{1/2,0}(z,\bar z) \rangle$ which was evaluated in \cite{KlebanSimmonsZiff06}. 

We used (\ref{QR}-\ref{QLR}) to construct the expression (\ref{ApproxSIStrip}) for $C(z)$ in the main text, but as we note in the introduction, the equivalent vertical quantity in this limit becomes
\be
C_v(z)=\frac{P_{T B}(z)}{\sqrt{P_T(z)P_B(z)\Pi_v}}=1\; ,
\ee
because $\Pi_v = 1$ while $P_T=P_{T B}+P_{T \bar B}$ and $P_B=P_{T B}+P_{\bar T B}$ both equal $P_{T B}$ since $\Pi_h=0$.  This is what we expect for crossing of a narrow rectangle.

\bibliography{SimmonsZiffKleban08v5}

\vfill\eject

\end{document}